%% file: main.tex
\documentclass[12pt,aps,nofootinbib,preprintnumbers,superscriptaddress,numberedappendix]{openjournal}

\usepackage{natbib}
\usepackage{newtxtext,newtxmath}
\usepackage[T1]{fontenc}
\usepackage{lmodern}  

\usepackage[colorlinks=true, allcolors=blue]{hyperref}
\DeclareRobustCommand{\VAN}[3]{#2}
\let\VANthebibliography\thebibliography
\def\thebibliography{\DeclareRobustCommand{\VAN}[3]{##3}\VANthebibliography}

\usepackage{graphicx}
\usepackage{amsmath}
\usepackage{microtype}
\usepackage{xspace}
\usepackage{makecell}
\usepackage{needspace}

\newlength{\colwidth}
\newlength{\figsizehalf}
\newlength{\figsizefull}
\shorttitle{Explore Gaia RVS stellar spectra with ML}
\shortauthors{Eilat et al.}

\begin{document}

\title{Exploration of groups and outliers in Gaia RVS stellar spectra with metric learning}
\author{
Yarden Eilat Bloch$^{1,\dag}$, 
Dovi Poznanski$^{1,2,3,4,*}$, 
Nick L.~J. Cox$^5$,\\ 
Emmanuel Bernhard$^5$, 
Iain McDonald$^6$,
Manuela Rauch$^7$,
and Albert Zijlstra$^6$ \\[0.5em]}

\affiliation{
$^1$School of Physics and Astronomy, Tel-Aviv University, Tel-Aviv 69978, Israel\\
$^2$Cahill Center for Astrophysics, California Institute of Technology, Pasadena CA 91125, USA\\
$^3$Kavli Institute for Particle Astrophysics \& Cosmology, 452 Lomita Mall, Stanford University, Stanford, CA 94305, USA\\
$^4$Department of Physics, Stanford University, 382 Via Pueblo Mall, Stanford, CA 94305, USA\\
$^5$ACRI-ST, Centre d’Etudes et de Recherche de Grasse (CERGA), 10 Av. Nicolas Copernic, 06130 Grasse, France\\
$^6$Jodrell Bank Centre for Astrophysics, University of Manchester, Oxford Road, Manchester M13 9PL, UK\\
$^7$ Know Center GmbH, 34 Sandgasse, 8010 Graz, Austria
}

\thanks{$\dag$~\href{mailto:yardy151@gmail.com}{yardy151@gmail.com}}
\thanks{$*$~\href{mailto:dovi@tau.ac.il}{dovi@tau.ac.il}}

\begin{abstract}
\noindent 
\input abstract
\end{abstract}

\maketitle




\input 01_introduction
\input 02_gaiaRVS
\input 03_trainingRF

\input 04_umapIntro
\input 05_thePlatform

\input 06_results
\input 07_discussion

\section*{Acknowledgments}
This project has received funding from the European Union’s Horizon 2020 research and innovation programme under grant agreement No 101004214 - \href{https://explore-platform.eu}{EXPLORE}. This research was also funded in part by the Koret Foundation, the Kavli Institute for Particle Astrophysics and Cosmology at Stanford University, and by grant NSF PHY-2309135 to the Kavli Institute for Theoretical Physics (KITP). Y.E.B. and D.P. acknowledge support from Israel Science Foundation (ISF) grant 541/17, and by grant 2018017 from the United States-Israel Binational Science Foundation (BSF).
This work has made use of data from the European Space Agency (ESA) mission
 \href{https://www.cosmos.esa.int/gaia}{\it Gaia}, processed by the {\it Gaia}
\href{https://www.cosmos.esa.int/web/gaia/dpac/consortium}{Data Processing and Analysis Consortium (DPAC)}. Funding for the DPAC has been provided by national institutions, in particular the institutions
participating in the {\it Gaia} Multilateral Agreement.

\bibliography{all}
\bibliographystyle{aasjournal}

\end{document}

%% file: abstract.tex
The Gaia mission is transforming our view of the Milky Way by providing distances towards a billion stars, and much more. The third data release (DR3) includes nearly a million spectra from its Radial Velocity Spectrometer (RVS). Identifying unexpected features in such vast datasets presents a significant challenge. It is impossible to visually inspect all of the spectra and difficult to analyze them in a comprehensive way. In order to supplement traditional analysis approaches, and in order to facilitate deeper insights from these spectra, we present a new dataset together with an interactive portal that applies established self-supervised metric learning techniques, dimensionality reduction, and anomaly detection, to allow researchers to visualize, analyze, and interact with the Gaia RVS spectra in straightforward but under-utilized manner. We process 312,295 Gaia DR3 RVS spectra (from 997,162 available spectra) by focusing on the high-quality subset with SNR $> 50$, which reduces noise-dominated artifacts in the anomaly ranking and makes the identified outliers more interpretable. We demonstrate a few example interactions with the dataset, examining groupings and the most unusual RVS spectra, according to our metric. This combination of methodology and public availability enables broader exploration, and may reveal yet-to-be-discovered stellar phenomena.

%% file: 01_introduction.tex
\section{Introduction}

The analysis of stellar spectra offers a critical window into astrophysical processes, revealing insights into stellar composition, kinematics, and evolution. The Gaia mission \citep{2016A&A...595A...1G,gaia_mission}, designed to produce the most precise 3D map of the Milky Way, has provided unprecedented data for over one billion stars, including detailed spectroscopic observations from the Radial Velocity Spectrometer (RVS) for about a million sources, as of the third data release (DR3; \citealt{2023A&A...674A...1G}). This kind of dataset also represents an immense potential for identifying unexpected populations and rare stellar phenomena.

Traditional analysis pipelines, such as GSPSPEC \citep{Recio_Blanco_2023}, are invaluable for the analysis and exploitation of such large data samples. However, by design they rely on detailed astrophysical models and are inherently limited by the parameter spaces these models encompass. By choice, these models are built to fit most objects well, which means that they will not fit some unknown minority of sources, some of which may be interesting. Such objects are often hard to find and can remain undetected or poorly characterized. 

 To overcome these limitations, researchers have increasingly applied machine-learning and representation-learning methods to large spectroscopic surveys. Examples include data-driven label methods such as \textit{The Cannon} \citep{Ness_2015}, machine-learning classification of interstellar features \citep{10.1093/mnras/stv977}, support-vector- and regression-based approaches developed for LAMOST (e.g., SLAM; \citealt{Zhang_2020}), and convolutional and deep-learning applications to surveys such as RAVE, Gaia-ESO, and others \citep[e.g.,][]{Guiglion2020,Ambrosch_2023}. Generative and adversarial models have also been used to recover and enhance astrophysical signal in low-S/N or degraded data \citep[e.g.,][]{10.1093/mnrasl/slx008,2307.06378}.

More recently, contrastive and multimodal self-supervised methods have been used to learn embedding spaces for astronomical spectra directly. In the stellar context, these include contrastive or CLIP-like models trained on Gaia XP and RVS spectra, or jointly with other modalities, which report physically structured latent spaces and useful downstream performance for similarity search, parameter inference, translation across survey products, and anomaly discovery \citep{buck_schwarz_2024,rizhko_bloom_2025,specclip_2025}. Related work has extended similar ideas to other astronomical spectral domains, for example by learning joint image-spectrum embeddings for galaxies and contrastive embeddings for IFU galaxy data \citep{parker_2024_astroclip,martinezsolaeche_2024}. In parallel, the broader metric-learning literature has proposed alternative data-driven distance constructions, including locally adaptive density-aware metrics and learned density-based distances \citep{kleiber_2024_laminar,sorrenson_2025}.

Building on previous applications of metric learning in astronomical spectroscopic datasets \citep{10.1093/mnras/stw3021,10.1093/mnras/sty348}, we extend these methods to the Gaia DR3 RVS dataset. In contrast to the recent neural contrastive and multimodal approaches above, our work uses a random-forest-based self-supervised pretext task to define similarity directly from the spectra themselves, without requiring paired modalities or a learned cross-instrument latent alignment. We performed the first large-scale self-supervised classification of Gaia RVS spectra and developed an interactive portal that enables broad community access to these insights.

Our methodology relies on Random Forests (RF; \citealt{2001MachL..45....5B}) trained on real and fake spectra to establish a meaningful metric space, allowing for the quantification of anomalies via a ``weirdness score.'' To further facilitate the exploration of high-dimensional spectral data, we use UMAP (Uniform Manifold Approximation and Projection; \citealt{mcinnes2018umap-software}), a dimensionality reduction technique that enables visualization of data structure in a two-dimensional space. This combination of RF-based metric learning and UMAP offers a useful framework for identifying outliers and gaining insights into the Gaia RVS dataset. 

Given the size of the dataset, and the numerous angles one might use for approaching it, even within our restricted framework, we do not attempt any sort of complete analysis in this work. Instead, as advocated and done previously for spectra and images of galaxies for example \citep{reis2019,Storey_Fisher_2021,Lochner_2021}, we present an interactive tool (and all the code to create it), for others to explore. This was done as part of the \href{https://www.explore-platform.eu/glance}{EXPLORE} project, whose aim is to facilitate efficient, user-friendly exploitation of data from astrophysics and planetary space missions and supporting ground-based surveys. The portal presented here enables users to visualize, filter, and investigate the Gaia RVS dataset in real time, democratizing access to large-scale spectroscopic analysis and potentially uncovering new stellar phenomena through community-driven exploration. 

This paper is structured as follows. In Section~\ref{Gaia-RVS-data}, we describe the Gaia DR3 RVS dataset and preprocessing steps, and in Section~\ref{s:methods} we detail our methodology. Section~\ref{s:platform} describes our data product and the interactive exploration tool we developed. In Section~\ref{sec:interactions} we demonstrate the utility of our tools and the potential for discovery by studying unusual groupings and the most anomalous spectra. We conclude with a discussion of implications and future directions in Section~\ref{s:discussion}.

%% file: 02_gaiaRVS.tex
\section{Input data}
\label{Gaia-RVS-data}

For this work we use combined and continuum-normalized RVS spectra obtained from Gaia DR3. These are primarily meant to provide crucial radial velocity measurements. The instrument is an integral-field spectrograph that operated in the near-infrared band ($845-872$~nm), with medium resolution ($\lambda / \Delta\lambda \sim 11500$). DR3 includes the spectra of 997,162 stars collected during the first 34 months of Gaia's nominal mission. These are the type AFGK stars with median per-bin signal-to-noise ratio (SNR)~$ > 20$ in the complete Gaia catalog \citep{gaia_dr3}. These spectra have been averaged over multiple (Average of ~20) rest-frame-aligned observations for increased SNR. Furthermore, the spectra have been normalized at the local pseudocontinuum and resampled to a wavelength bin width of 0.01~nm \citep{refId0}. We trim about 1~nm at each end of the supplied spectra to avoid missing values, resulting in a wavelength range of $847.5-868.5$~nm with 2100 fluxes per object.

In early stages of the analysis, we noticed that objects with low SNR tended to dominate the set of objects classified as anomalous. In order to focus on objects which are less ambiguous, and more rewarding to interpret, we excluded objects with an SNR (\textit{rv\_expected\_sig\_to\_noise} or \textit{rvs\_spec\_sig\_to\_noise}) lower than 50. This results in an overall sample size of 312,295 out of the 997,162 published RVS spectra. Future Gaia releases, including DR4, are expected to provide higher-SNR RVS products and RVS time-series information, which could make a larger fraction of the spectra useful for similar analyses and enable time-domain extensions of this kind of exploration.

%% file: 03_trainingRF.tex
\section{Methods}
\label{s:methods}
Our purpose is to learn directly from the data with limited biases and preconceptions. By using Metric Learning, we compare objects to each other to generate a meaningful similarity metric. It is meaningful in that it allows one to find outliers and study the distribution of stellar spectra. As detailed below, we use two main algorithms in combination. First, a random forest classifier is trained on the data, using a pretext task to derive similarities. We then study these similarities in a lower dimensional space, using dimensionality reduction. Our methods are based on the works of \citet{10.1093/mnras/stw3021} and \citet{10.1093/mnras/sty348} who developed and applied these approaches to galactic and stellar spectra.

\begin{figure}[ht!]
\centering
\includegraphics[width=\figsizefull]{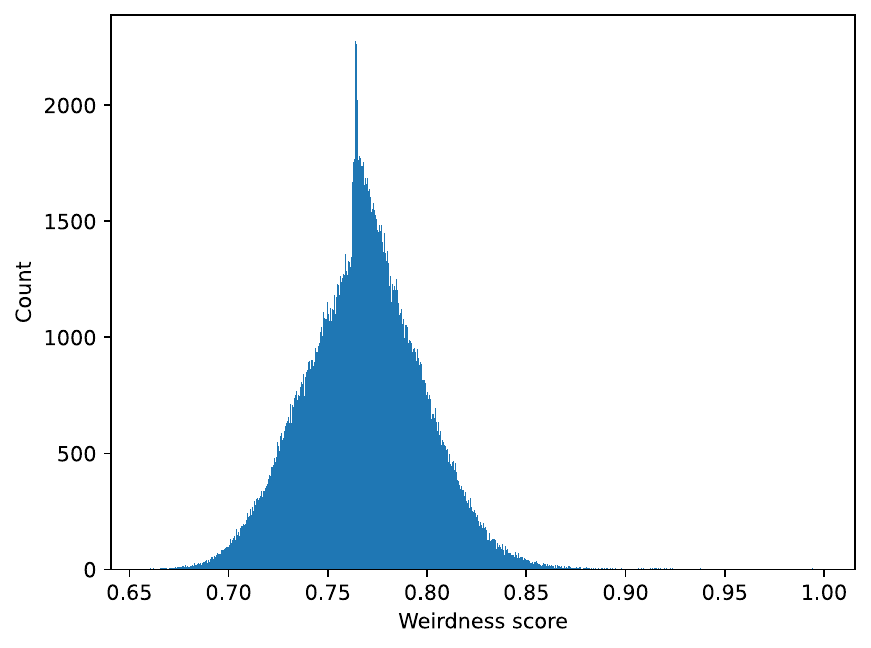}
\caption{Distribution of weirdness scores for the sample. The narrow spike near 0.76 is due to a group of extremely similar stars as discussed in section \ref{sec:weirdness_spike}}
\label{fig:weirdness_hist}
\end{figure}

\subsection{Random Forest Architecture}

As typically done in metric learning, we use a self-supervised pretext task to train a classifier that learns to focus on the features that distinguish spectra from random noise. A pretext task is an auxiliary learning problem whose solution requires the model to learn a useful similarity structure from the data. In this context, “self-supervised” means that the training labels are generated automatically from the data itself rather than from external annotations such as stellar types. The task provides a learning signal that forces the classifier to capture the intrinsic correlations present in real spectra.

We construct a sample of random fake spectra, where the flux value at each wavelength is drawn independently from the marginal distribution of the real spectra at that wavelength. This preserves the overall flux distribution at each wavelength but destroys the correlations between wavelengths that exist in real spectra. The training dataset therefore consists of both real and fake spectra, labeled respectively as “real” and “fake.” In the high-SNR configuration used for the final analysis, this pool contains 199,432 real spectra and 99,716 fake spectra. We first split this full pool 80/20 into training and test sets, reserving the 20\% test set for evaluation. For computational efficiency, the RF is then fit not on the entire training split, but on a random 50,000-example subset drawn from it (33,261 real and 16,739 fake examples). We find the main structure of the learned metric broadly insensitive to moderate changes in this training-subsample size. In the \textit{Scikit-learn} implementation used here, each tree is trained on a bootstrap resampling of this 50,000-object fit subset, while at each split the tree considers only a random subset of the wavelength bins.

By construction, the fake dataset lacks the short- and long-range correlations between different wavelengths that one sees in real spectra. In a real spectrum, values within an emission line are correlated, and there are many positive and negative correlations between various lines. The continuum of a spectrum is by definition correlated. Therefore, for the classifier to succeed in separating real from fake spectra, it must learn these correlations — effectively learning what makes a spectrum “real.”

The trained RF consists of an array of $N$ classification trees, each tree having multiple leaves labeled either "real" or "fake", and each trained using a subset of the spectra and a subset of the wavelengths. After training, we pass only the real spectra through the forest again. Each spectrum then lands in one terminal leaf per tree, and the combination of these leaf indices defines an $N$-dimensional embedding for that spectrum.

Spectra that are similar to each other tend to fall into the same terminal leaves across many trees, and therefore end up close to each other in the embedding space. A measure of similarity (or distance) can thus be defined for every pair of spectra as the fraction of trees in which they land in the same terminal leaf, referred to as the Hamming distance. Repeating this for all object pairs yields a pairwise distance matrix. Using the distance matrix, we calculate a “weirdness score” for each object, defined as the average distance to all other objects in the dataset. Objects with large scores are outliers that can be studied further to determine what made them unusual.
This definition is a deliberate choice: weirdness is a global outlier statistic by construction, because every object is compared to the full sample. Other valid definitions exist, including explicitly local scores such as Local Outlier Factor (LOF), where the scale of outlierness is controlled by the neighborhood size $k$. In this work we keep weirdness as the primary score for consistency with the RF-distance framework and for direct interpretability across the full catalog; we discuss the implications of this choice in Section \ref{sec:weirdest_objects} and in the conclusion.

\begin{figure*}[hp!]
  \centering
  \begin{tabular}{@{}cc@{}}
    \includegraphics[width=0.49\textwidth]{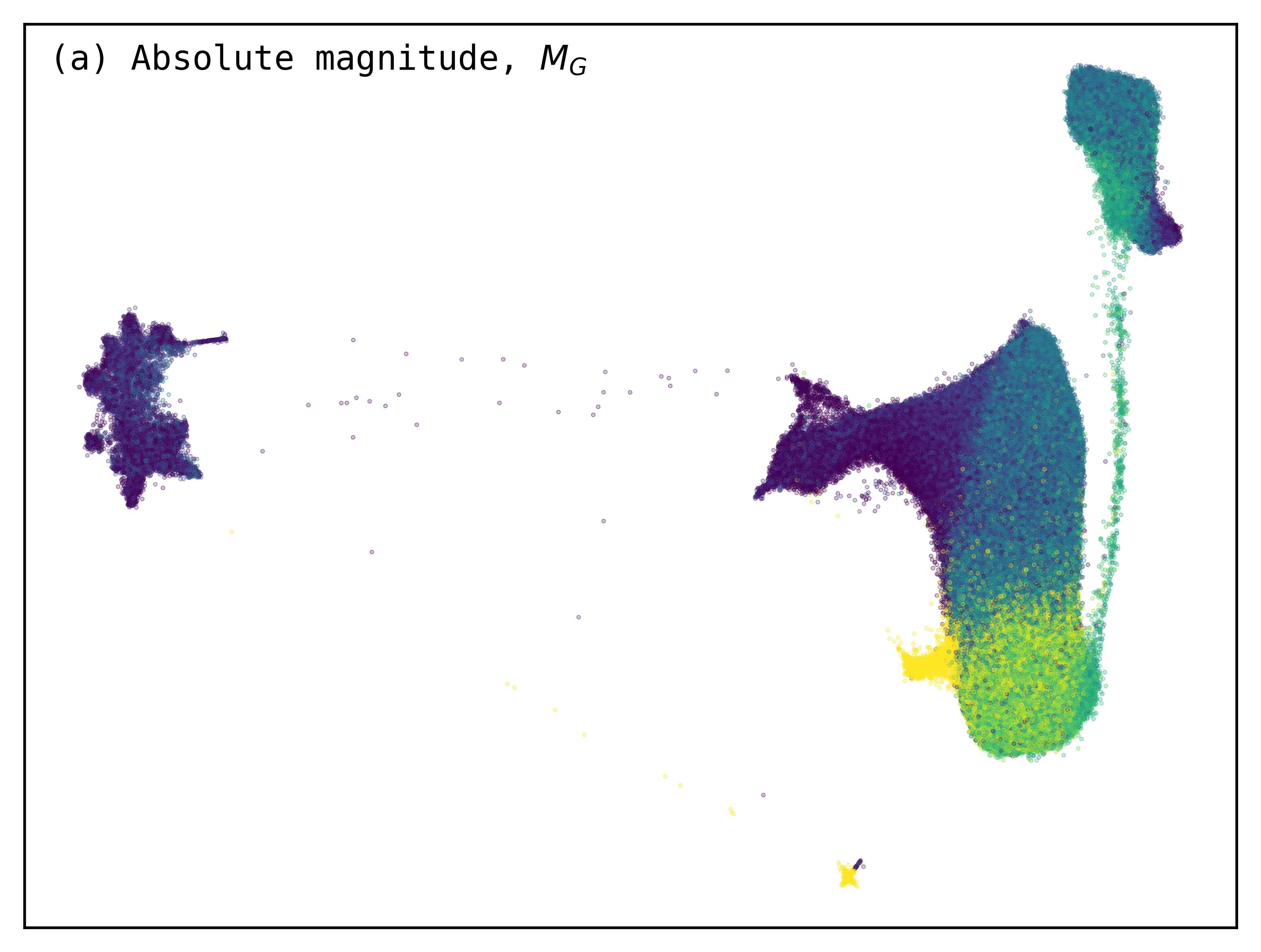} &
    \includegraphics[width=0.49\textwidth]{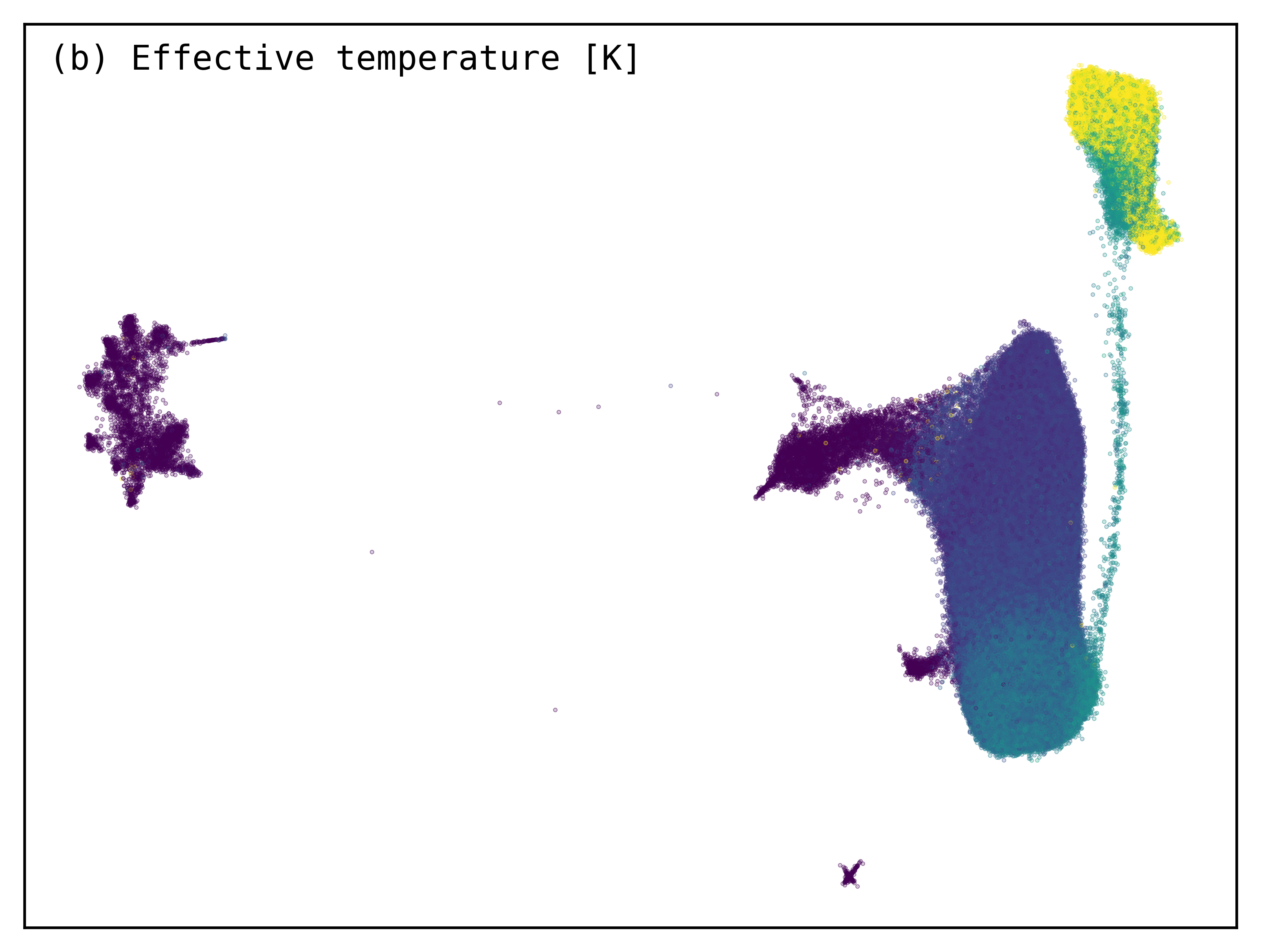} \\
    
    \includegraphics[width=0.49\textwidth]{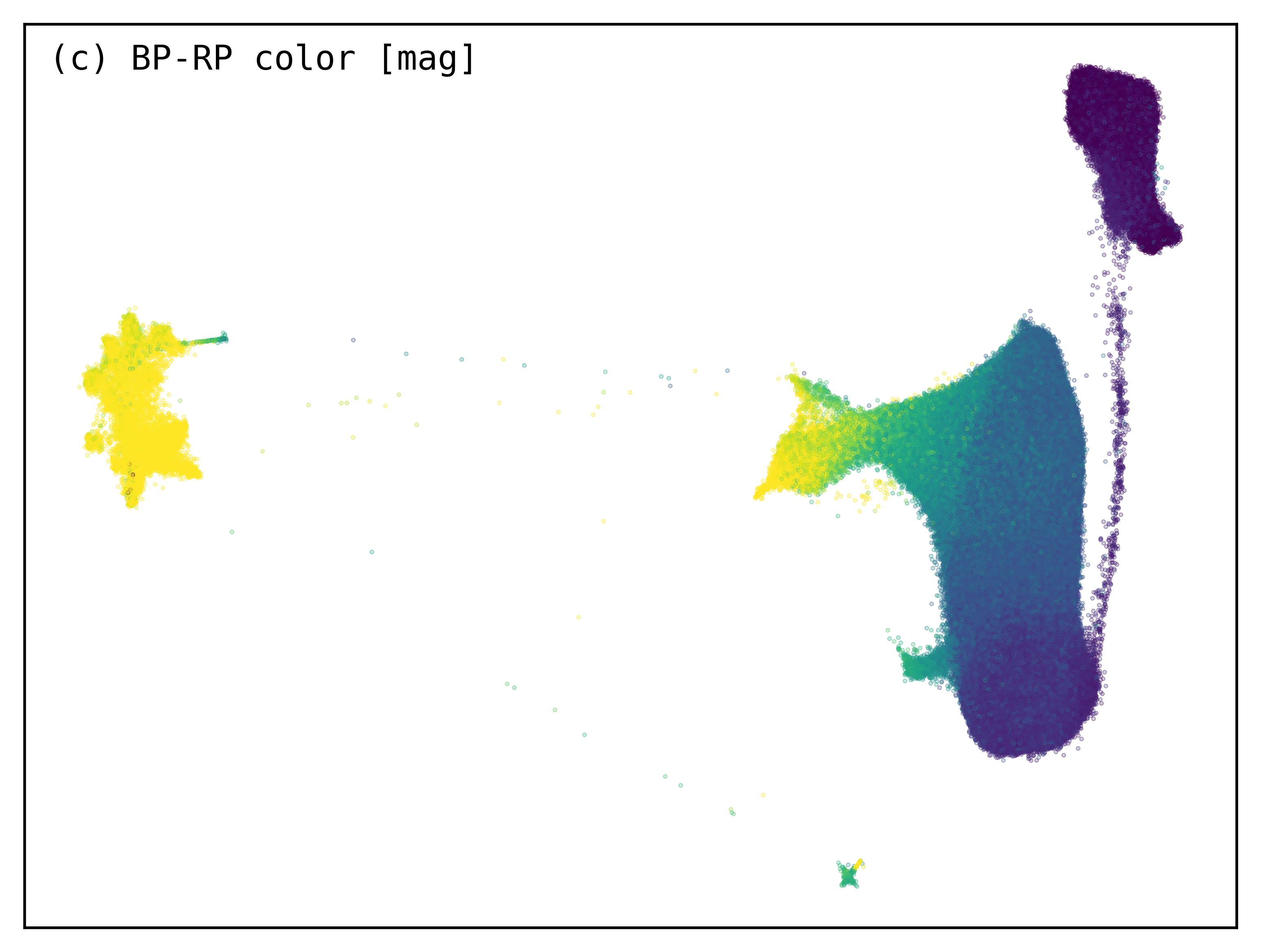} &
    \includegraphics[width=0.49\textwidth]{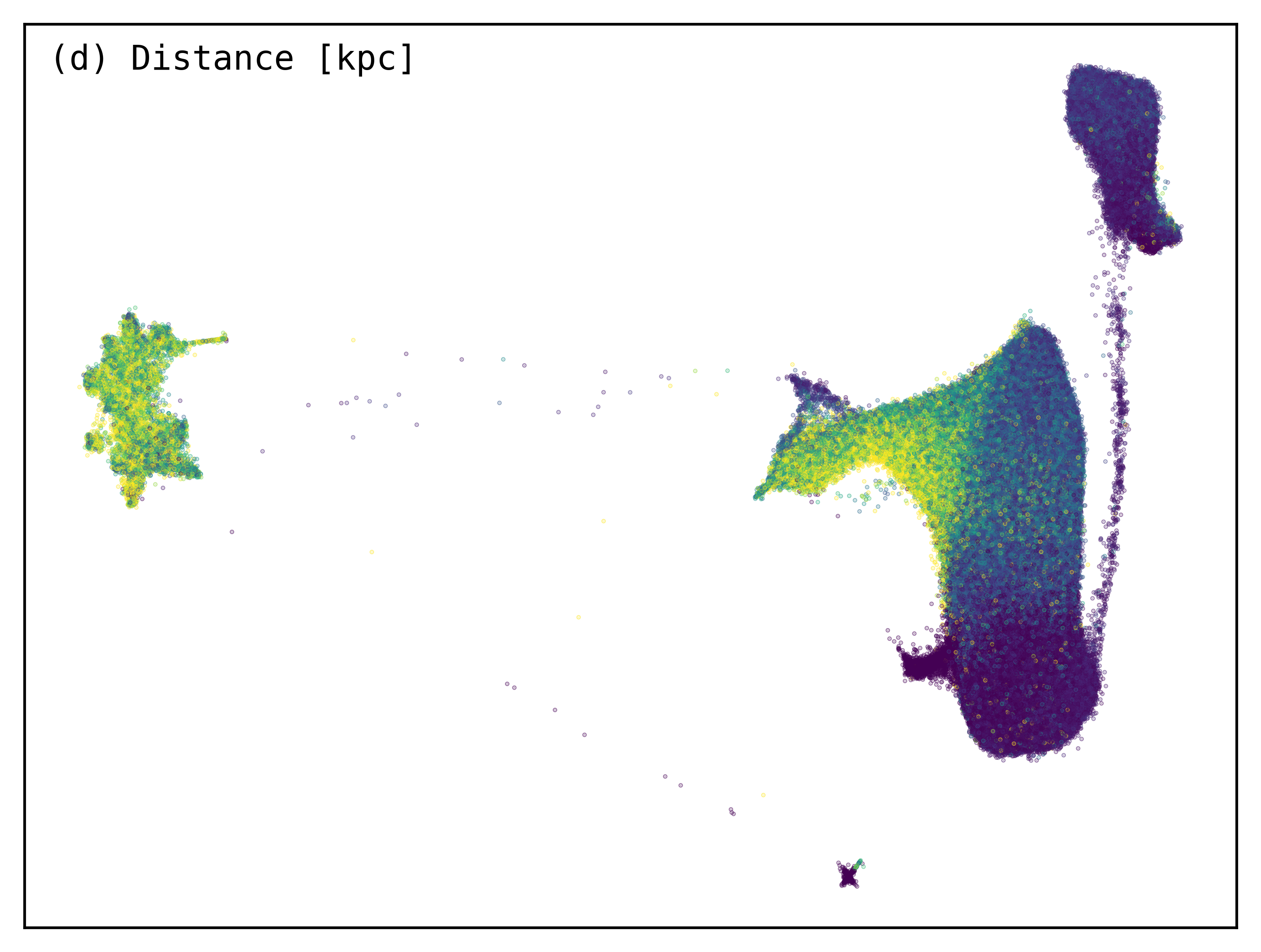} \\

    \includegraphics[width=0.49\textwidth]{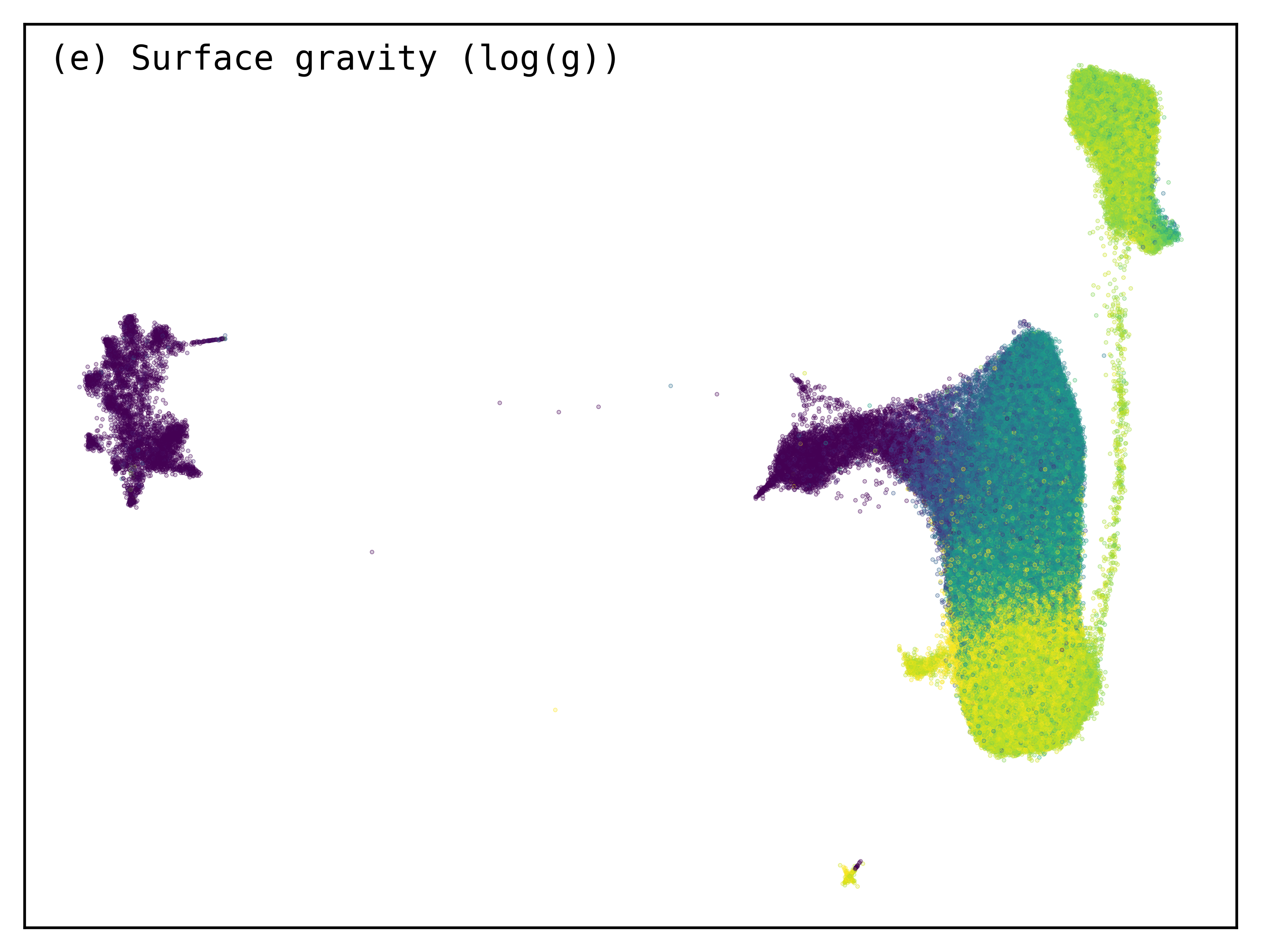} &
    \includegraphics[width=0.49\textwidth]{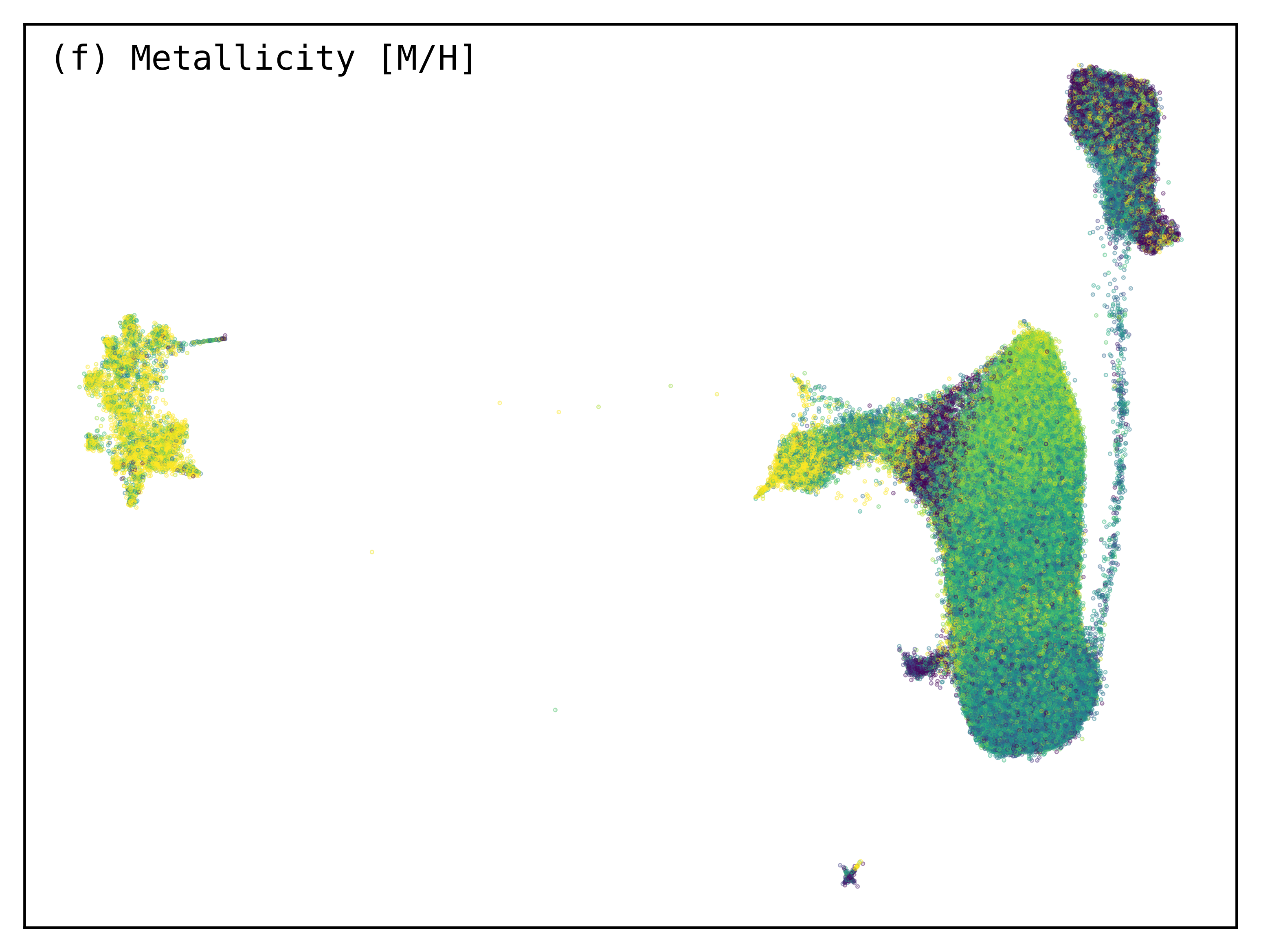}
  \end{tabular}

  \caption{UMAP projection of our distance matrix. Each point on the map represents a star, where spectrally-similar objects are projected closer together. The axes do not have a direct physical interpretation since the transformation is non linear. The colors in every panel are:  (a) Absolute G-band magnitude, (b) Photometry-derived effective temperature, (c) BP-RP color, (d) distance, (e) surface gravity, and (f) metallicity. The values of these properties are derived by the GSPPHOT pipeline. Stars with missing values in the plotted quantity are excluded from that panel only; these omissions affect only a small minority of the sample and do not alter the overall map morphology.}
  \label{fig:umap_projections}
\end{figure*}

We use the \textit{Scikit-learn} implementation of Random Forest (RF) in Python \citep{scikit-learn}, which offers control over a range of hyper-parameters. Unlike in a supervised learning setting, we do not really have an objective function to optimize, except for the successful classification in the pretext classification task. However, achieving high classification accuracy was possible with a broad range of hyper-parameter combinations. So to gauge the utility of an embedding we instead rely on specific indicators of a meaningful metric.

Primarily, we expect the population to be mostly `normal'  with relatively few rarer objects. This should be reflected in a weirdness score distribution that is relatively low for the vast majority of objects, but with a tail that reaches higher scores. Additionally, when looking at the clustering properties of the data (mostly visually after dimensionality reduction) we expect it to lie on a low-dimensional and relatively continuous manifold, rather than many well separated clusters. This is due to the fact that there are few if any hard spectral subdivisions between classes of stars. Instead we usually expect a mostly continuous distribution in observed properties, tracing the mostly continuous distributions in Teff, log(g), metallicity, and other properties.

During the training process, we mainly experimented with three key hyperparameters of the random forest classifier: the number of trees ($n\_estimators$), the minimum number of samples required to split an internal node ($min\_samples\_split$), and the number of features to consider when seeking the best split ($max\_features$).

The number of trees ($n\_estimators$) directly impacts both the computational efficiency and the quality of the classifier. After conducting experiments with increasing numbers of trees, we identify that going beyond 500 trees does not significantly alter the resulting low-dimensional structure. At this point, the classifier achieves a perfect accuracy of 1. This suggests that adding more trees beyond this threshold is unnecessary. We therefore set the number of trees to be $500$ resulting in a $500$-dimensional embedding.

We find that using 500 trees together with $min\_samples\_split = 30$ strikes the right balance and encourages the emergence of a cohesive global structure in the subsequent UMAP visualization. In the saved high-SNR model used for the final figures and outlier analysis, the per-split feature sampling follows the RF default ($max\_features=\sqrt{2100}$, i.e. 45 spectral bins considered at each split), rather than a fixed one-fifth of the spectrum. These choices are obviously arbitrary and not unique. Importantly, the feature subsampling does not mean that each tree sees only a fixed 45-dimensional projection of the data. Rather, different random subsets of wavelengths are proposed at different splits across many trees, so the ensemble can still capture correlations across the full spectral range. The main effect of reducing the training-set size or making the per-split feature subsampling too aggressive is therefore not a simple ``curse of dimensionality'' failure, but a noisier estimate of which wavelength correlations are robustly informative. In practice, smaller training sets mainly increase the variance of the learned metric and make rare populations less stably represented, whereas the dominant large-scale organization and the highest-weirdness objects remain qualitatively similar over reasonable choices.

Figure \ref{fig:weirdness_hist} shows the distribution of weirdness scores for the entire dataset. Its shape is arbitrary, and results from random sampling and from our particular choice of hyper-parameters. Our only qualitative requirement for that distribution was that it should contain a relatively small fraction of high-weirdness objects, so that these scores actually represent a level of oddity. The peculiarly narrow peak in the distribution, close to a weirdness value of 0.76, is due to a group of very similar stars, all falling in the exact same leaves, as we show in Section \ref{sec:weirdness_spike}.

%% file: 04_umapIntro.tex
\subsection{Dimensionality Reduction}
\label{umap}
High-dimensional data are difficult to analyze and even harder to visualize, yet this is our purpose here. The RF step produces a 500D embedding. We use UMAP to reduce it down to a more manageable 2D. 

UMAP operates by constructing a low-dimensional representation of the data points while preserving the relationships (distances) between nearest neighbors. It achieves this by approximating a topological manifold on which the data lies to create a faithful representation in lower-dimensional space \citep{2018arXivUMAP}.

\begin{figure*}[ht!]
  \centering
  \begin{tabular}{cc}
    \includegraphics[height=0.35\textwidth]{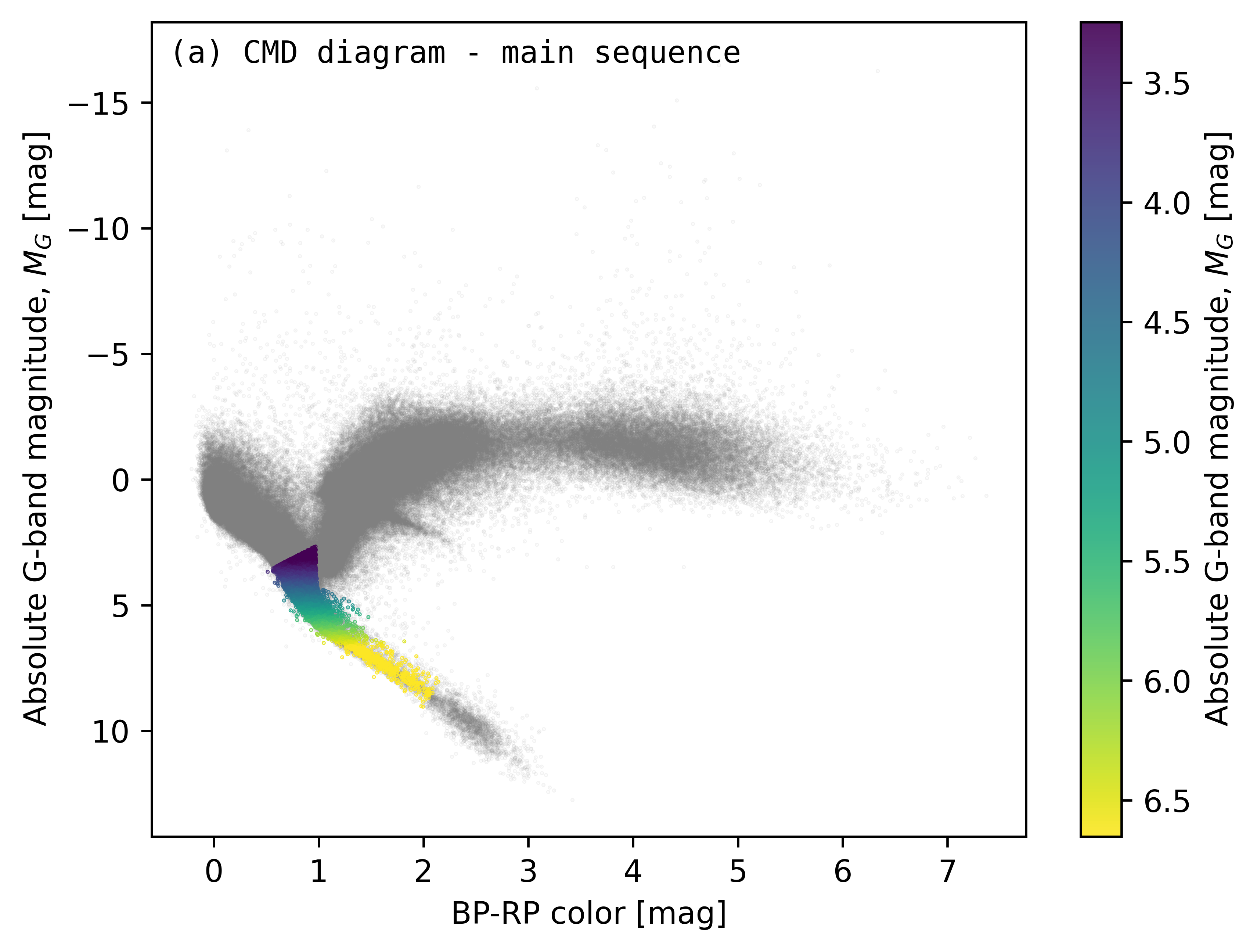}
    \includegraphics[height=0.35\textwidth]{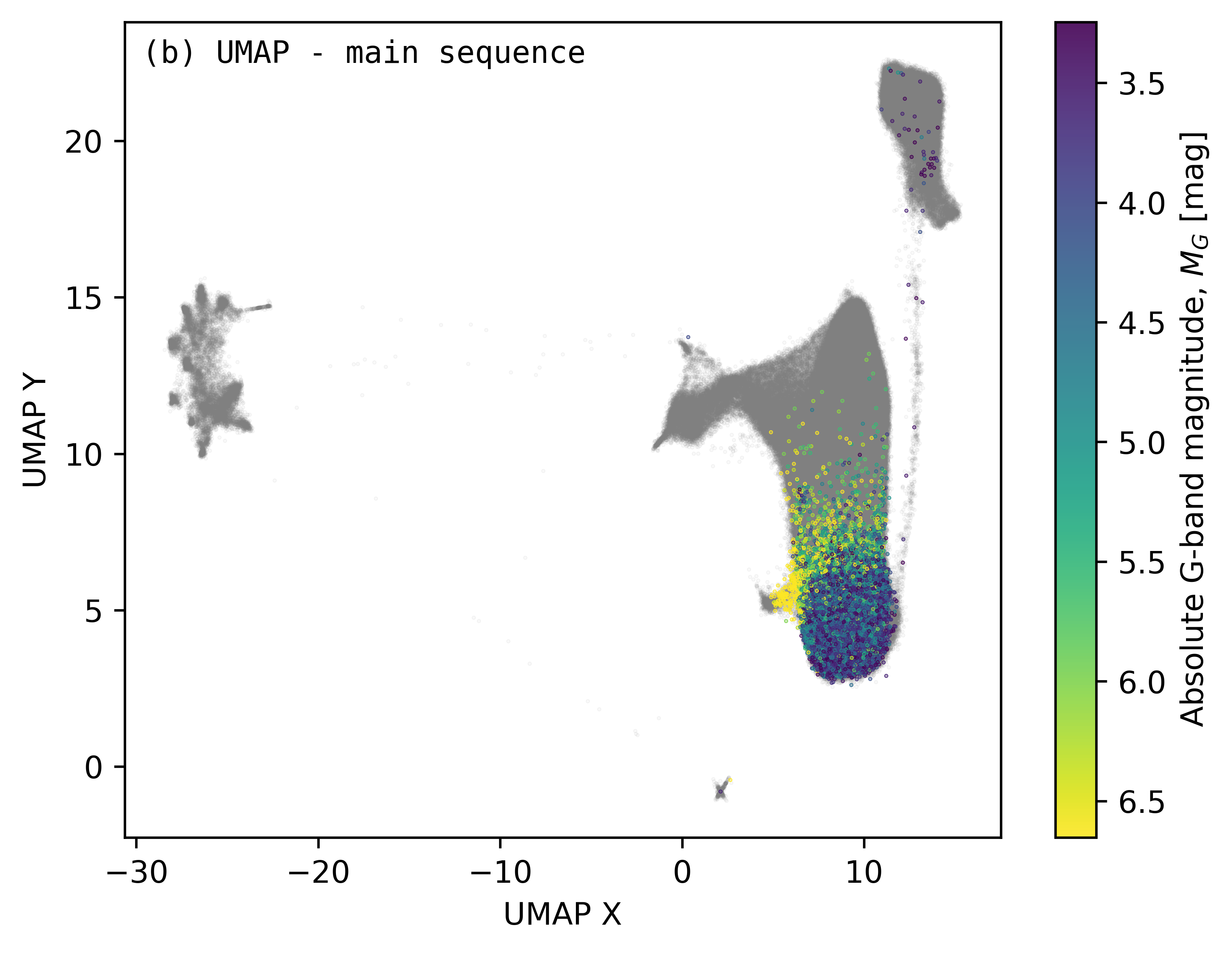}
    \end{tabular}
    \caption{Main-sequence stars highlighted in the CMD (with absolute G-band magnitude, $M_G$, on the vertical axis) and in UMAP. The highlighted sample is the middle main-sequence CMD locus used here as an illustrative subset for comparing CMD-space and UMAP-space neighborhoods.}
    \label{fig:main_seq_hr_umap}
\end{figure*}

Once we trained the RF, every object can be represented by the terminal leaf it ends up in, in every one of the 500 trees. Along every one of these dimensions, two objects are either in the same leaf, or not. The distance between two objects is the fraction of the dimensions that are different (Hamming distance).
We use the \textit{umap-learn} python package implementation of UMAP \citep{mcinnes2018umap-software}. We set 'hamming' as the distance metric for learning the UMAP representation, and since we are interested in a 2D visualization, $n\_components$ was set to 2. The hyper-parameters $min\_dist$ and $num\_neighbors$ were  set to 0.5 and 25 (respectively) as these seemed to strike a good balance between local and global structures in the resulting representation. For computational efficiency, UMAP was fit on a random subset of 50,000 objects, and the remaining objects were then mapped with the learned transform step (i.e., no additional manifold-learning optimization was performed when adding the rest of the sample). This implies that very rare populations that are poorly represented in the fit subset can be projected into the existing geometry without necessarily forming a distinct isolated island in 2D, even if they are globally unusual in the full 500D space. Our statement that results are broadly insensitive refers to the stability of the main large-scale structures under reasonable subset and hyperparameter choices, not to an expectation that every extreme outlier morphology is invariant. Still, any single UMAP configuration emphasizes only a limited range of scales, so it cannot display all structure in the 500D RF embedding equally well. In particular, using smaller values of $min\_dist$ and $num\_neighbors$ can reveal finer local substructure, while larger values emphasize broader global organization.

Every spectrum is now represented by a point in a 2D abstract space reduced from the 2100 spectral fluxes, or the more instructive 500 dimensions of the RF similarity. To confirm that indeed proximity in this mapping is a tracer of spectral similarity, we can color the points using various known properties for these stars. In Figure \ref{fig:umap_projections} we plot multiple times the same UMAP projection, but colored by (left to right, top to bottom) absolute magnitude, effective temperature, photometric colour, distance, surface gravity and metallicity. When one of these derived quantities is unavailable, that star is omitted only from the corresponding color-coded panel; this same per-quantity masking policy is used consistently across the analogous diagnostic plots in the paper, and the number of such omissions is small relative to the full sample. Clearly the mapping that was learned with no supervision solely from the RVS spectra is tracing the physical properties of the stars in a non-linear fashion, with the x-axis roughly corresponding to (from left-to-right) increase in effective temperature, decrease in BP-RP color, and increase in surface gravity.

It is instructive to see how the UMAP space translates to the space of the Color-Magnitude Diagram (CMD), where we have a good understanding of the stellar zoo. In Figure \ref{fig:main_seq_hr_umap} we can see that the middle part of the main sequence mostly maps to a single region in UMAP. However, as we show in Section \ref{sec:interactions}, other stars get split into multiple clusters.  

%% file: 05_thePlatform.tex
\begin{figure*}[ht!]
	\centering
	\includegraphics[width=0.90\textwidth]{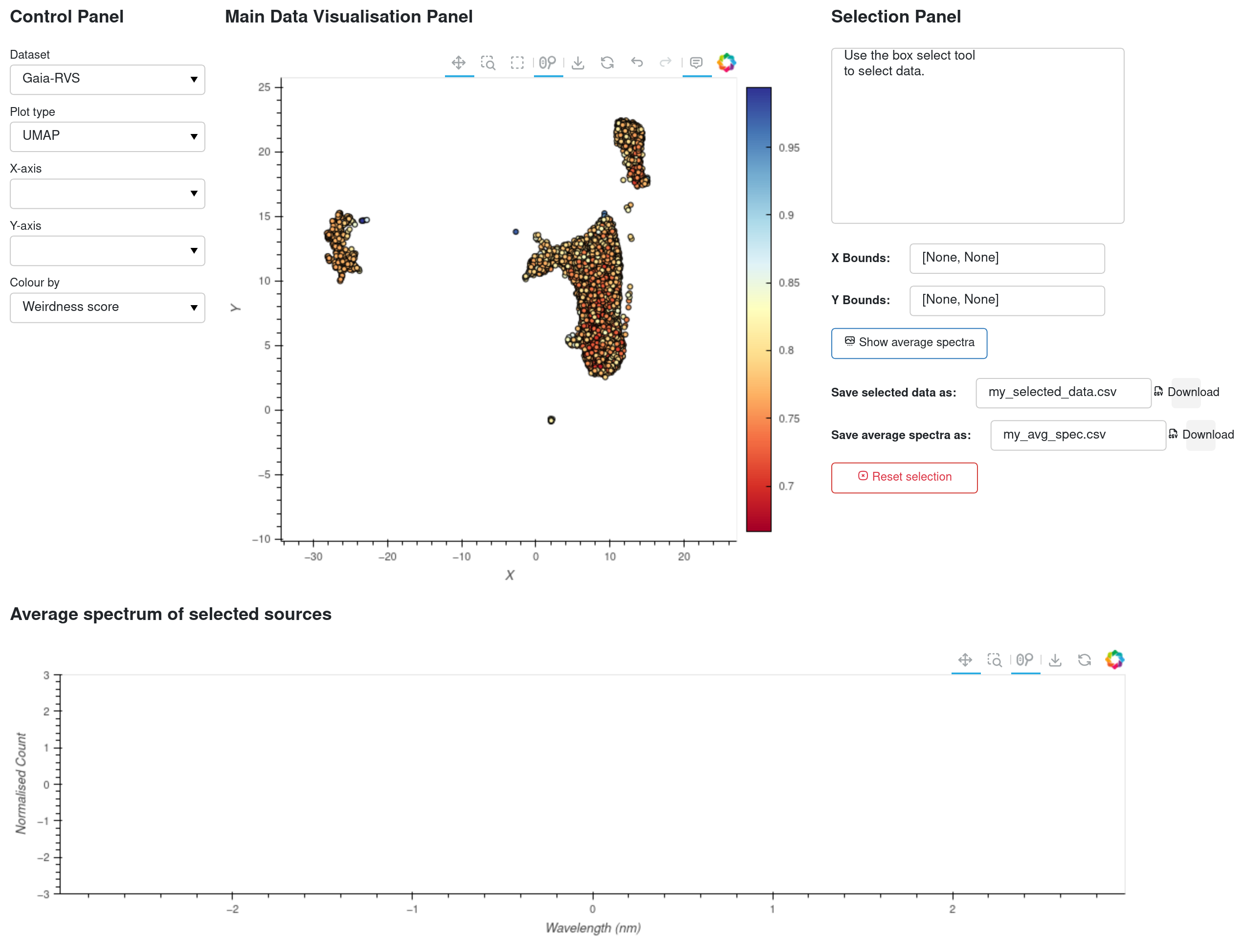}
	\caption{The user interface of the \href{http://sdisco.tau.ac.il/sdisco/sda}{S-Disco exploration tool}. The main figure is located in the center, plotting the database objects. Drop-down menus allow for control of the X-Y axes and color of the plotted objects. The figure includes buttons for zooming, panning, resetting, saving, and selecting objects. When objects are selected, they will appear in the list on the right, once plots are updated. The list can be saved with a dedicated button. The spectrum panel also supports single-object inspection: selecting one source (from the selection panel or directly on the scatter plot) and updating plots displays that object's spectrum.}
	\label{fig:main_page}
\end{figure*}

\section{The interactive tool}
\label{s:platform}

This analysis and product are part of EXPLORE, a project aimed to facilitate efficient, user-friendly exploitation of data from astrophysics and planetary space missions and supporting ground-based surveys. Six novel Scientific Data Applications have been created that relate to three thematic areas: galactic science, stellar characterisation, and lunar observation.

The web app discussed here is intentionally simple, featuring a single webpage with dropdown menus for selecting various plot types and coloring options to visualize the data (Figure \ref{fig:main_page}). Users can choose the plot type from the "Plot Type" dropdown menu, which includes the following options:
\begin{itemize}\itemsep-3pt
	\item \textbf{UMAP}: 2D representation of the object's embeddings.
	\item \textbf{Galactic Plane}: Object locations in Galactic X and Y coordinates.
	\item \textbf{Galactic Side-View}: Object locations in Galactic X and Z coordinates.
	\item \textbf{CM Diagram}: Color-magnitude diagram.
\end{itemize}

Object colors can be chosen from the "Color by:" dropdown menu, including other available parameters from the Gaia database (such as distance, effective temperature, metallicity, etc.) and the RF derived weirdness score.

The main figure displays the data as a scatter plot, where each Gaia object is represented as a colored point, with a color bar on the right. Hovering over a point reveals the Gaia ID and color value of the object. The main figure also supports zooming, panning, object selection using the lasso tool, and saving the figure via icons on the right.

After selecting objects, the following is triggered:
\begin{enumerate}\itemsep-3pt
	\item The selected object IDs populate a list on the right of the main figure, which can be saved to CSV by pressing the "Save selected data as" button.
	\item If "Show average spectrum" is selected, the median spectra and 95\% confidence interval of the selected objects are displayed in the lower panel.
\end{enumerate}
In particular, when the selection contains a single source, this panel provides direct single-spectrum visualization for that object.

The S-Disco tool is publicly available at \url{http://sdisco.tau.ac.il/sdisco/sda}. The tool can also be run locally using the source code at \url{https://github.com/explore-platform/s-disco}. The underlying dataset is available from \href{https://doi.org/10.5281/zenodo.16043466}{Zenodo (10.5281/zenodo.16043466)}.
Advanced users may use the provided code and files to perform more complex interactions not supported by the user interface, or recreate the analysis with different choices of hyper-parameters.

In the following section we will present various interactions that demonstrate the capabilities and explore some aspects of the RVS sample.

%% file: 06_results.tex
\section{Discussion of example interactions}\label{sec:interactions}

\Needspace{12\baselineskip}
\subsection{UMAP groups}
Once the objects are mapped to 2D there appears to be some clustering, or grouping of stars, and we explore here whether these are physical, instrumental, or in any way revealing. 
\Needspace{10\baselineskip}
\subsubsection{Secondary main sequence}
\begin{figure*}
    \centering
    \includegraphics[width=0.48\textwidth]{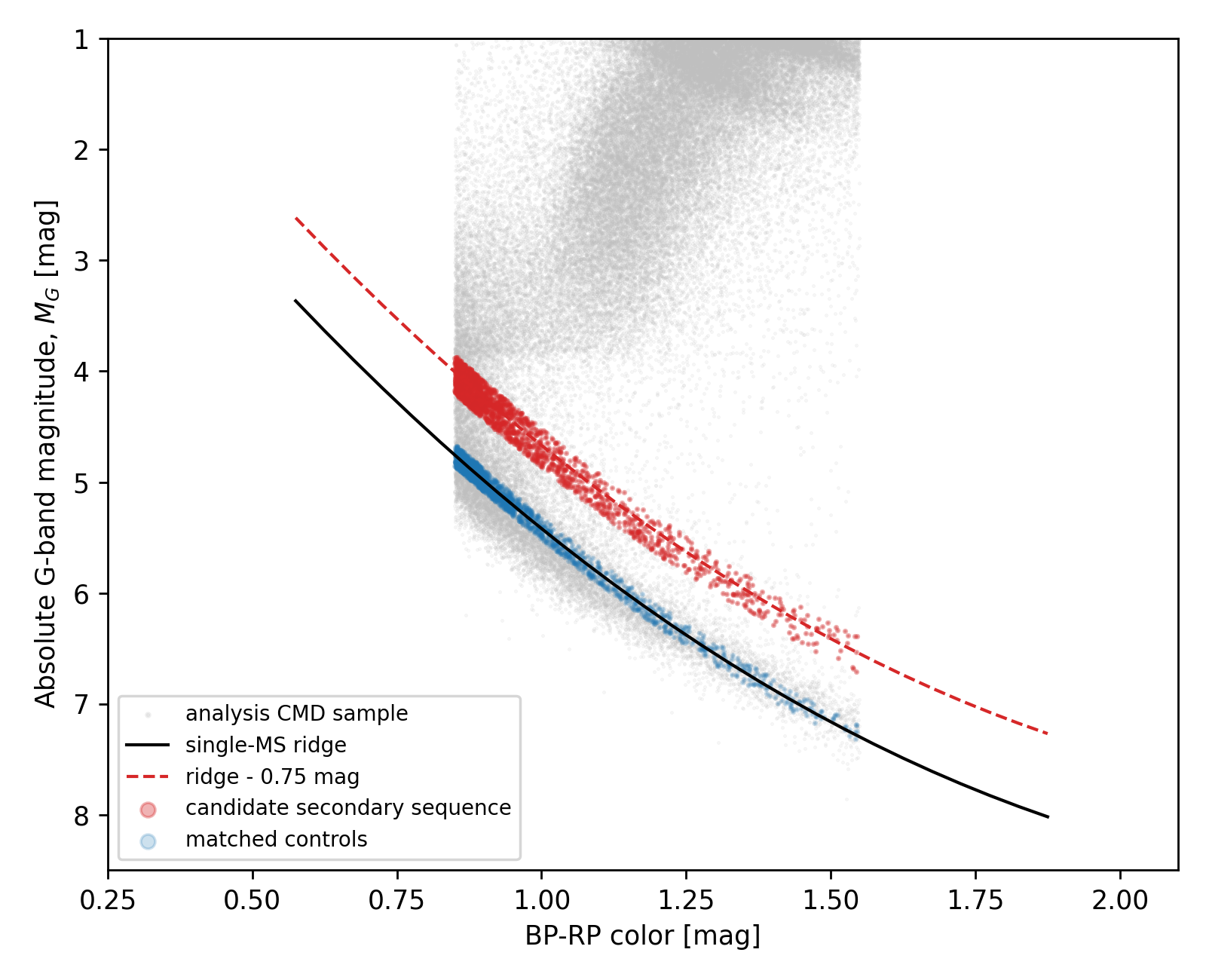}
    \hfill
    \includegraphics[width=0.48\textwidth]{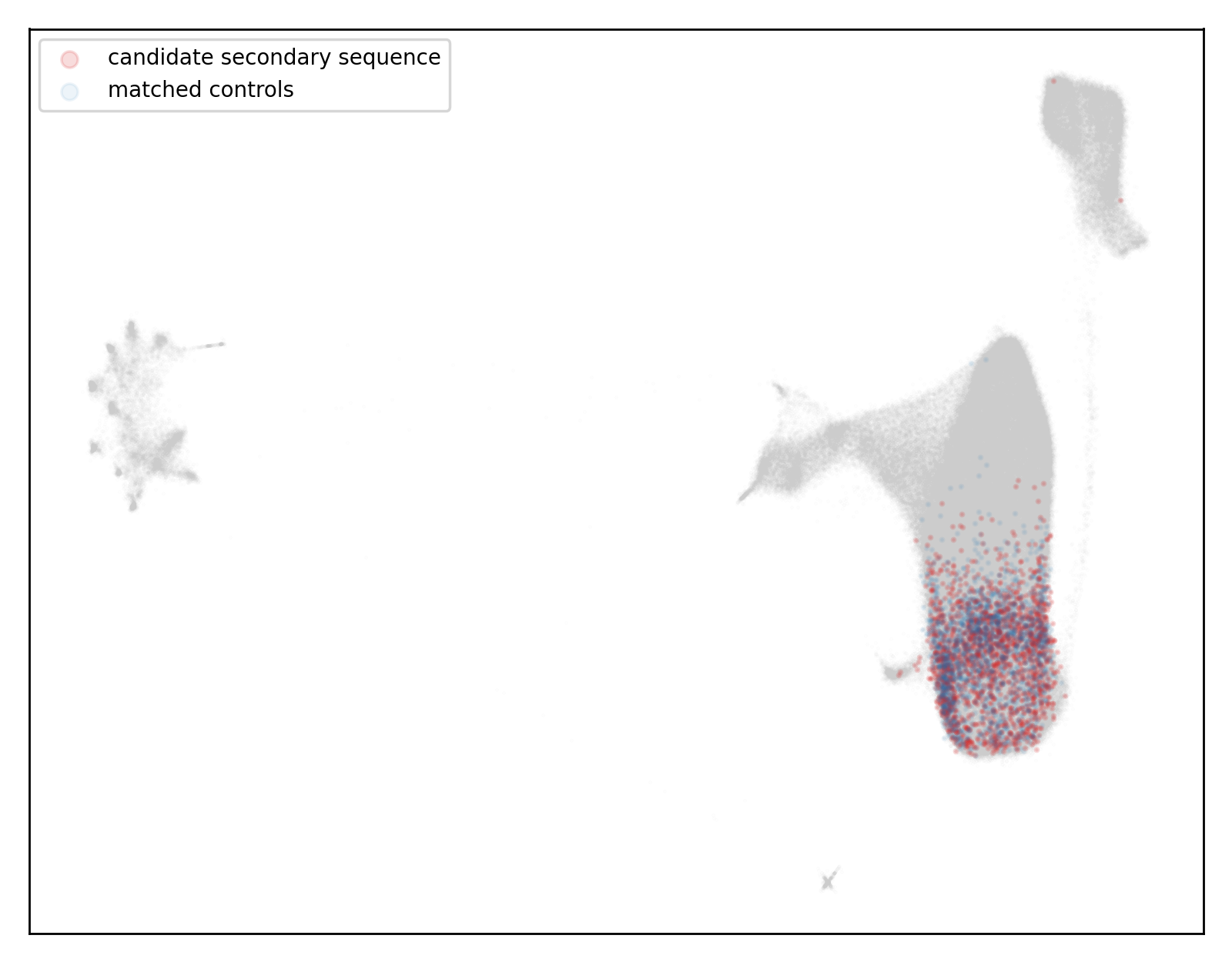}
    \caption{Secondary-main-sequence diagnostic selection. Left: observed Gaia CMD with the fitted single-star ridge, the ridge shifted brighter by 0.75 mag, the candidate photometric secondary-sequence sample, and the color-matched ridge controls. The single-star ridge is fitted to the 65th percentile of the main-sequence anchor sample in each color bin, and the selection is restricted to the mid-main-sequence color range to avoid the upper-main-sequence turnoff shoulder. Right: the same secondary-sequence candidates and controls shown in the UMAP projection; the two samples mostly overlap in the broad main-sequence region rather than forming a distinct UMAP island.}
    \label{fig:secondary_ms_cmd_umap}
\end{figure*}

\begin{figure*}
    \centering
    \includegraphics[width=0.48\textwidth]{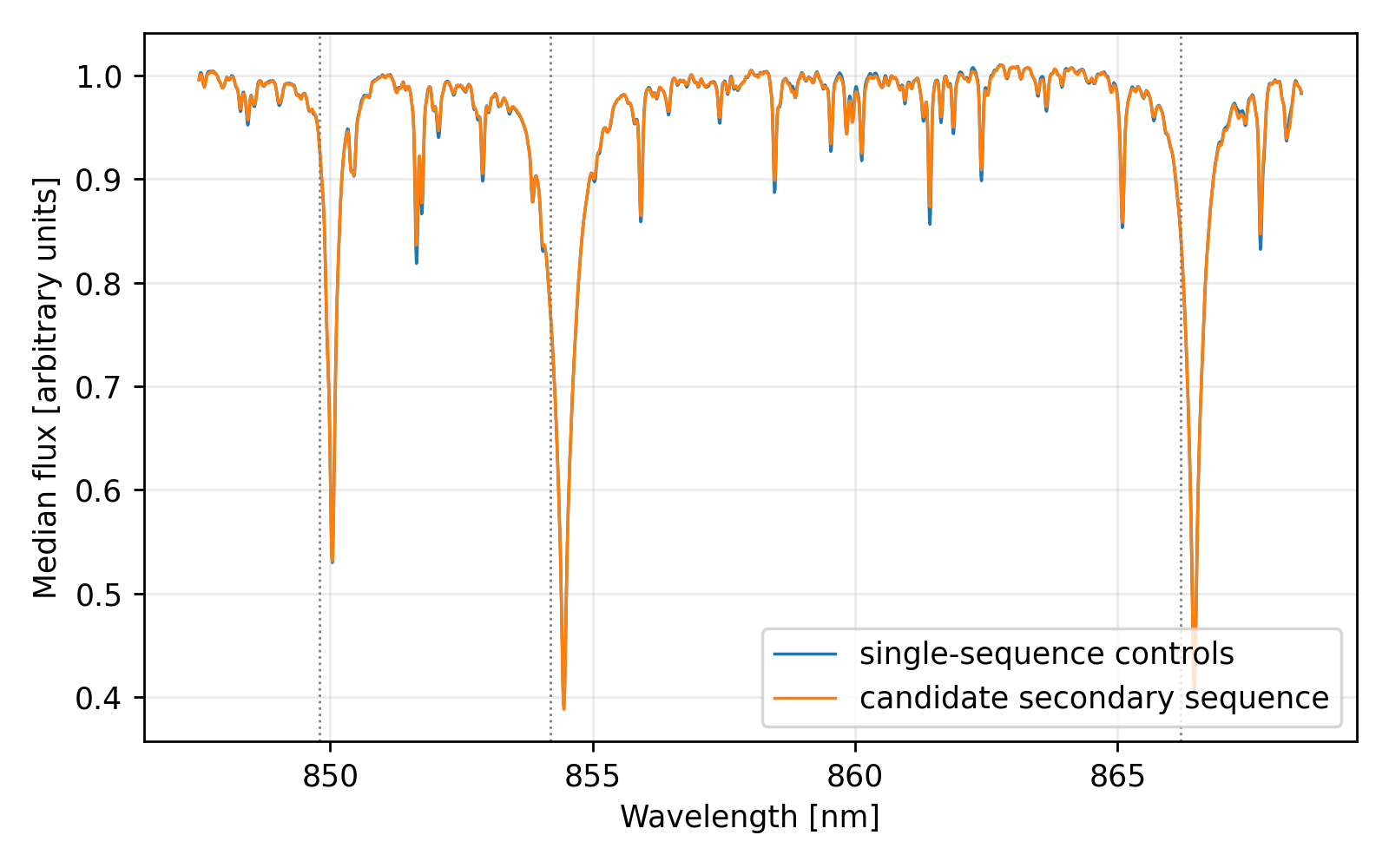}
    \hfill
    \includegraphics[width=0.48\textwidth]{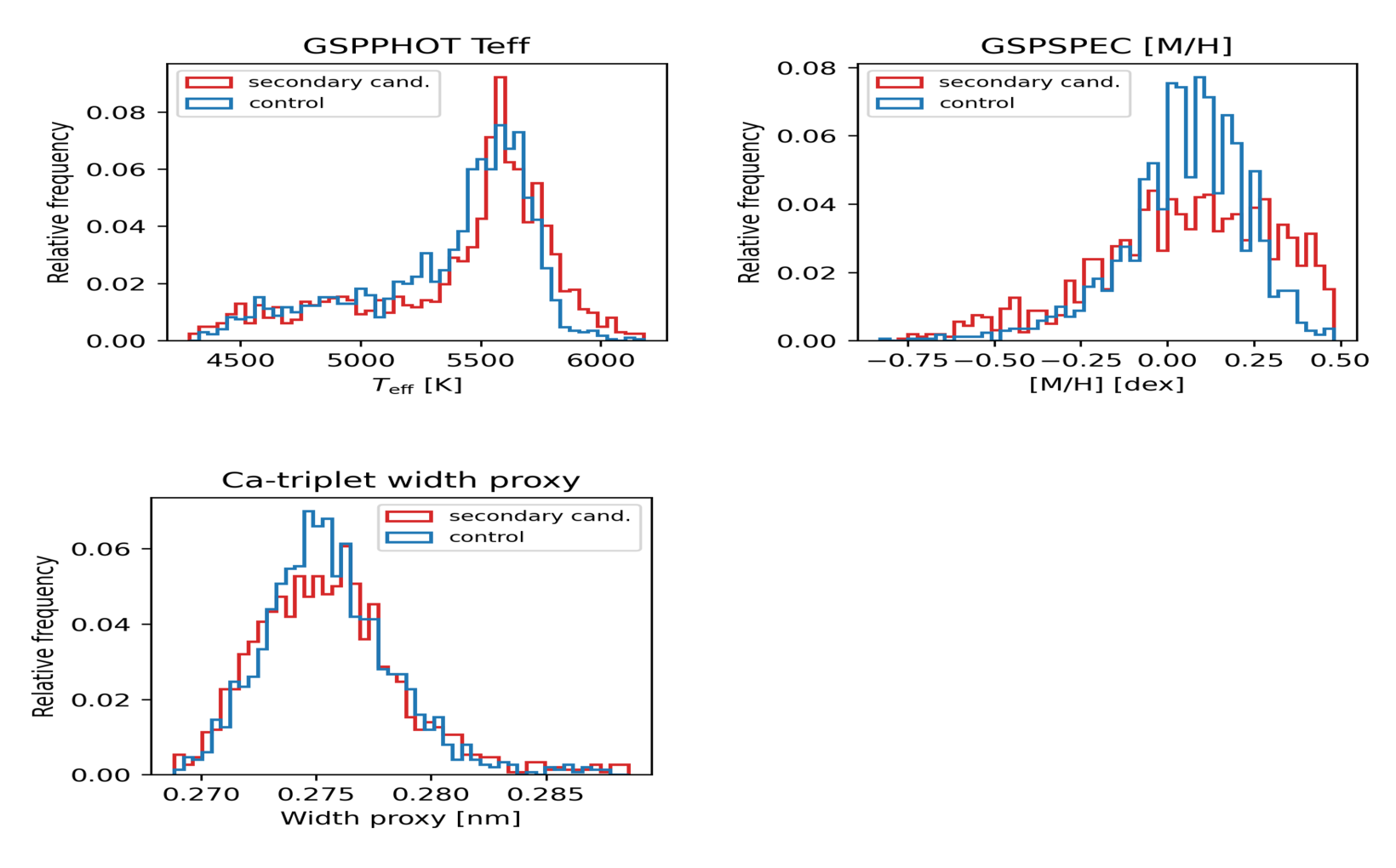}
    \caption{Spectral and label comparison for the candidate secondary main sequence. Left: median RVS spectra of the secondary-sequence candidates and the matched single-sequence controls, showing no visually resolved double-line structure in the median comparison. Right: diagnostic label and line-shape distributions; the effective-temperature and metallicity distributions are similar, and the second-moment Ca-triplet width proxy does not show a clear broadening excess for the secondary-sequence candidates.}
    \label{fig:secondary_ms_spectra_params}
\end{figure*}

The Gaia CMD also shows a secondary main sequence above the single-star locus, as expected for an unresolved-binary population \citep{HurleyTout1998}. To check whether this photometric sequence has an obvious RVS signature, we fitted the main-sequence ridge using the 65th percentile of the existing main-sequence anchor sample and selected a cleaner mid-main-sequence sample over $0.85\leq BP-RP\leq1.55$ lying $0.55$--$0.90$ mag above that ridge (Figure \ref{fig:secondary_ms_cmd_umap}). This gives 1,859 candidate photometric secondary-sequence stars, which we compared to 1,859 color-matched ridge stars. The two samples have similar effective-temperature distributions (median GSPSPEC $T_{\rm eff}\simeq 5630$ K for the secondary sequence and $\simeq 5520$ K for the ridge controls) and similar GSPSPEC metallicities (median $[M/H]\simeq 0.07$ and $0.06$, respectively; Figure \ref{fig:secondary_ms_spectra_params}). The main label difference is a modestly lower GSPSPEC $\log g$ for the secondary-sequence candidates (median 4.17 versus 4.26), which is expected for stars selected to be brighter at fixed color and is not by itself a spectroscopic binary diagnostic. We do not see a distinct secondary-sequence island in UMAP; instead these stars overlap the same broad main-sequence region as the ridge controls. We also checked representative spectra, median spectra, equivalent-width proxies, and a simple second-moment Ca-triplet width proxy for matched subsamples, finding no clear excess broadening or visually separated double-line structure in the secondary-sequence candidates. Thus, within the resolution and labels used here, the secondary CMD sequence is mainly a photometric feature rather than a population that our RVS-based metric separates strongly from comparable main-sequence stars.

\Needspace{10\baselineskip}
\subsubsection{Hot stars}
\begin{figure*}
  \centering
    \includegraphics[width=\textwidth]{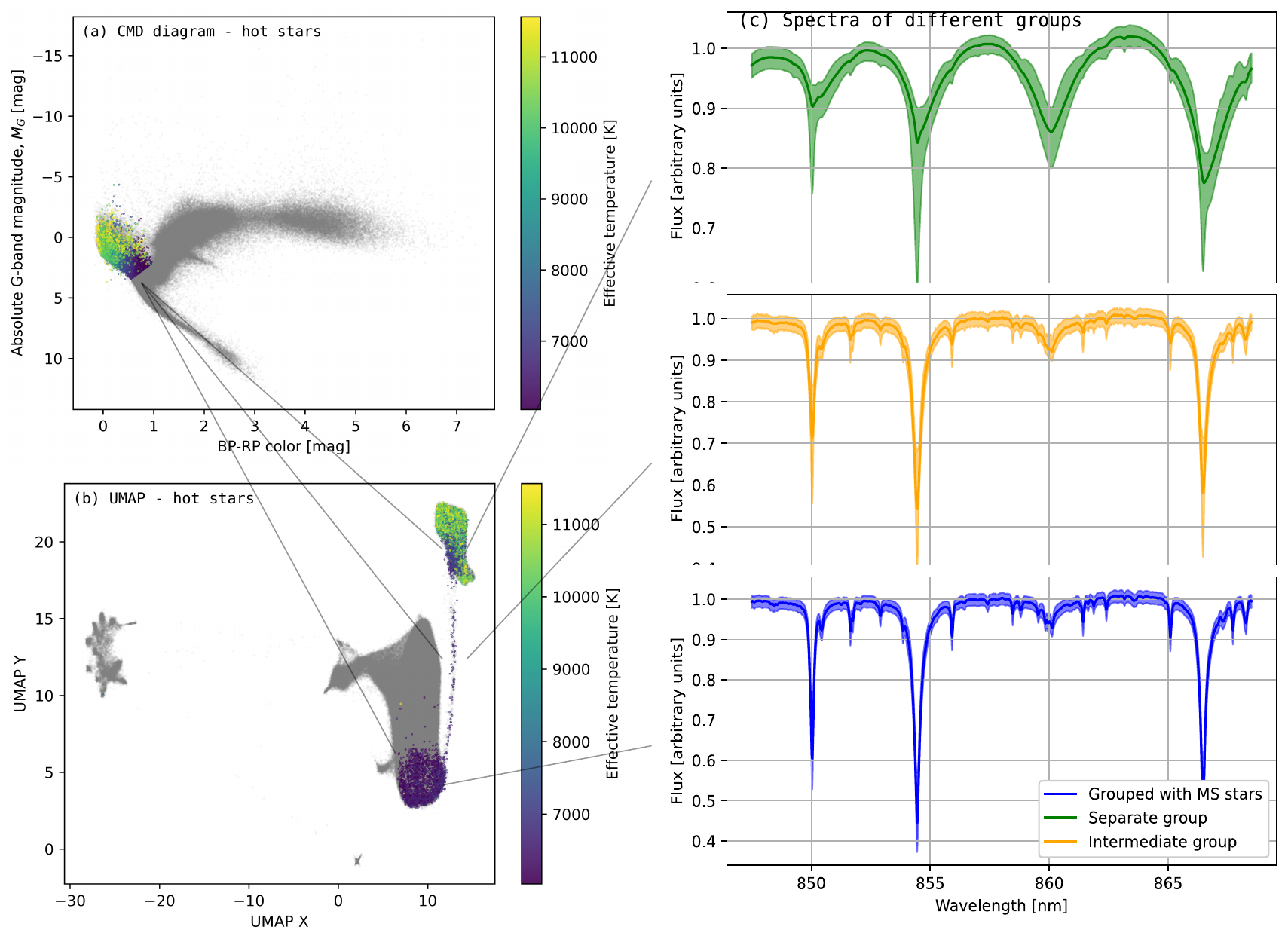}
    \caption{Hot stars highlighted in CMD and UMAP together with their different spectra. The CMD selection uses absolute G-band magnitude ($M_G<3$) and color ($BP-RP<1$).}
    \label{fig:blue_giant_hr_umap}
\end{figure*}

We select stars with approximately $M_G<3$ and $BP-RP<1$, as shown on the CMD in  Figure \ref{fig:blue_giant_hr_umap}a. These stars, which we refer to here as Hot stars, appear split into two distinct areas in UMAP, mostly along the vertical axis, as seen in Figure \ref{fig:blue_giant_hr_umap}b. The majority of them, about 30,000, are split to the top right, while about 8000 are not easily separable in this space from the bulk of the MS stars. There is also a small intermediate group of a few hundred stars that connect the two clusters.

\begin{figure}[!htbp]
    \centering
    \includegraphics[width=0.48\linewidth]{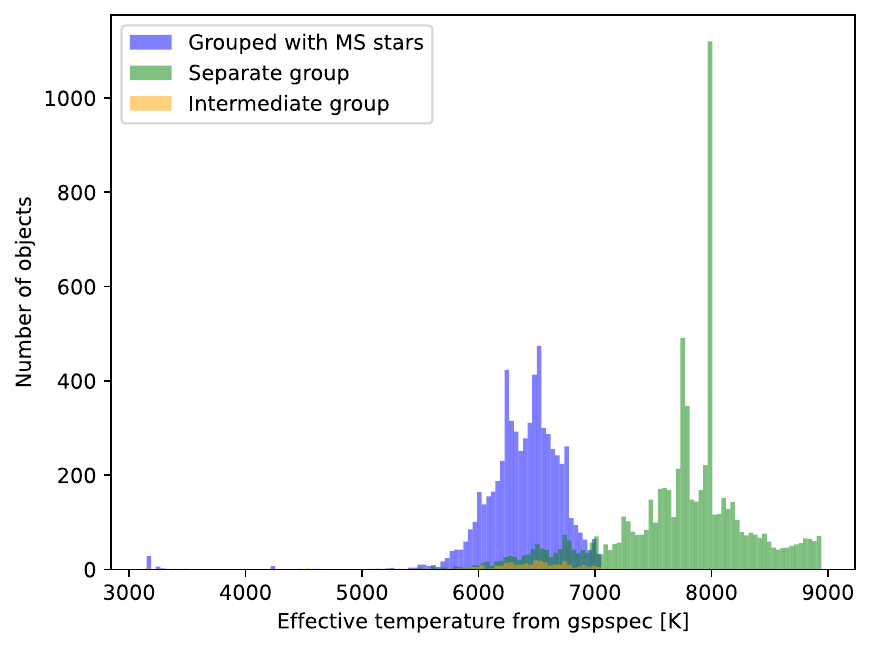}
    \hfill
    \includegraphics[width=0.48\linewidth]{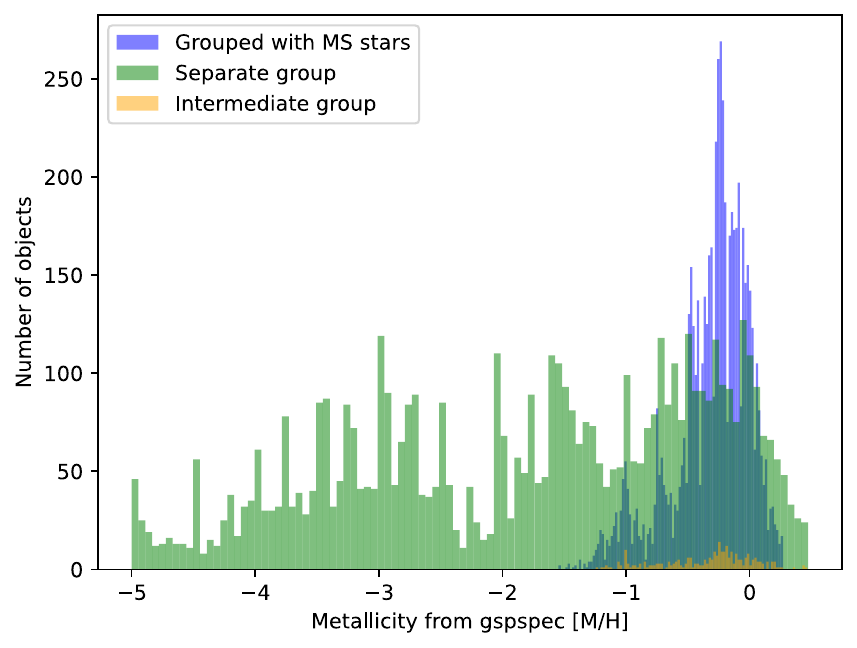}
    \caption{Distributions of $T_{\rm eff}$ (left) and metallicity (right) of the Hot-star groups obtained from the GSPSPEC pipeline. The metallicity panel should be interpreted with caution for the hottest Paschen-dominated spectra, where GSPSPEC metallicities are expected to be less reliable.}
    \label{fig:blue_giant_teff_mh_hist}
\end{figure}

\begin{figure}
    \centering
    \includegraphics[width=\figsizefull]{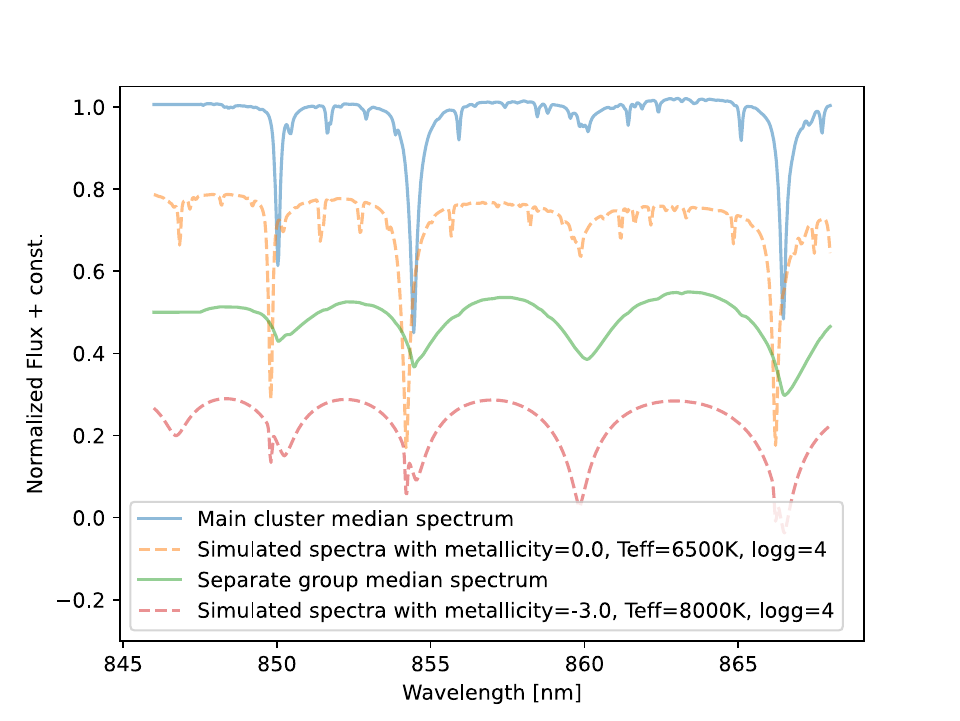}
    \caption{Simulated spectra from \citep{Prieto}, with properties chosen to visually match the two groups of Hot stars.  The temperatures are similar to what we find in GSPSPEC.}
   \label{fig:sim_spec_blue_giants}   
\end{figure}

We analyze the RVS spectra associated with each of these groups, and we show them in Figure \ref{fig:blue_giant_hr_umap}c. Clearly, the objects in the disjoint group have broader, shallower, and fewer absorption lines, and the separation in UMAP seems sound given the spectral dissimilarity. Analysis of the properties derived by the GSPSPEC pipeline for these groups reveals that the groups differ mainly in effective temperature and metallicity (Figure \ref{fig:blue_giant_teff_mh_hist}). Note the bimodal distribution in effective temperature, in alignment with the split into two distinct groups in the UMAP presentation with very few intermediate objects. However, we do not interpret the very low metallicities in the hottest subgroup as a literal young metal-poor population. In hot Gaia RVS spectra the Hydrogen Paschen lines dominate this wavelength range and lie close to the Ca II triplet, which complicates the interpretation of line-driven parameters in this regime \citep{Blomme2023HotRV}. More generally, Gaia DR3 treats hot stars with a dedicated parameter-estimation module and explicitly identifies regimes in which the reported parameters should be used with caution \citep{Fouesneau2023ApsisII}. We therefore use the metallicity histogram here only as a qualitative diagnostic, not as firm evidence for a physically metal-poor component. This is further confirmed by visually comparing the median spectra of each of the two Hot-star groups to simulated spectra from \citet{Prieto}, as shown in Figure \ref{fig:sim_spec_blue_giants}.

\subsubsection{Red dwarfs}
Similarly to the Hot stars, we define Red dwarfs loosely, referring to the area in the CMD corresponding to a G-band absolute magnitude more than 8 and BP-RP color greater than 2. These stars as well, when mapped with UMAP, separate into two distinct groups. The first group, closely associated with the main sequence stars, comprises around 1500 objects, while the separate group contains roughly 500 objects.

\begin{figure*}
  \centering
    \includegraphics[width=\textwidth]{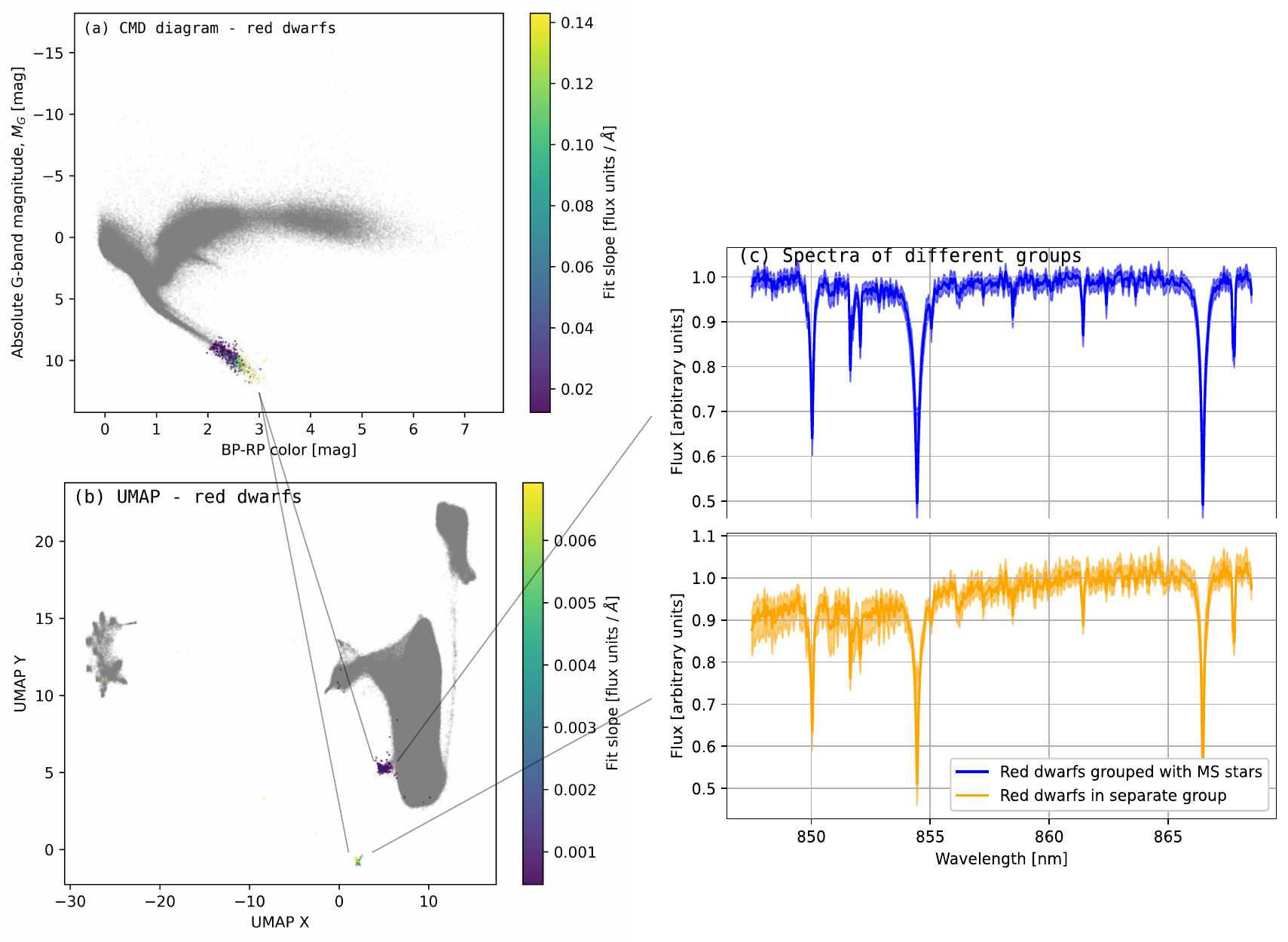}
    \caption{Red dwarfs highlighted in CMD and UMAP together with their different spectra. The CMD selection uses absolute G-band magnitude ($M_G>8$) and color ($BP-RP>2$).}
    \label{fig:red_dwarf_hr_umap}
\end{figure*}

\begin{figure}[ht]
    \centering
    \includegraphics[width=0.48\linewidth]{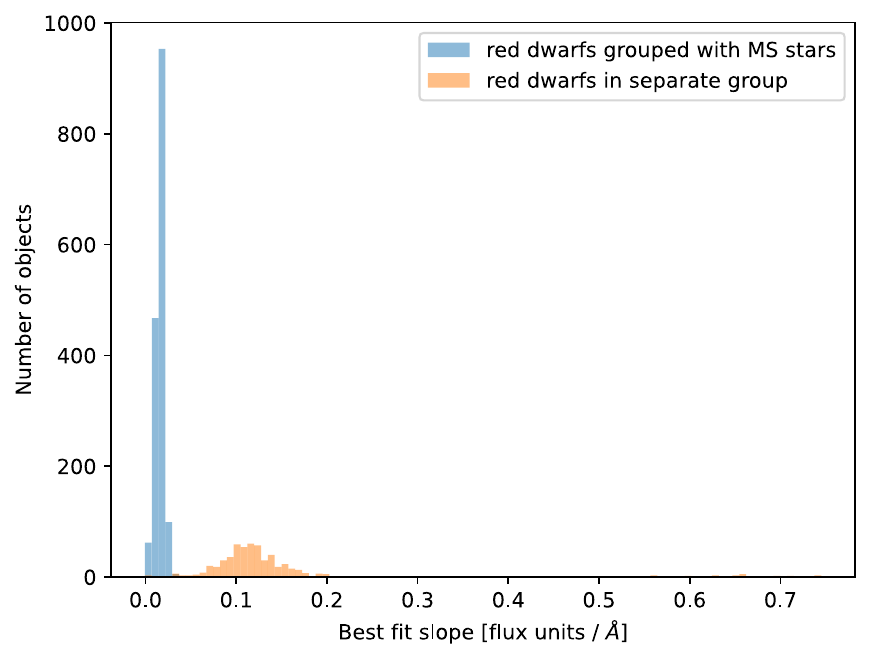}
    \hfill
    \includegraphics[width=0.48\linewidth]{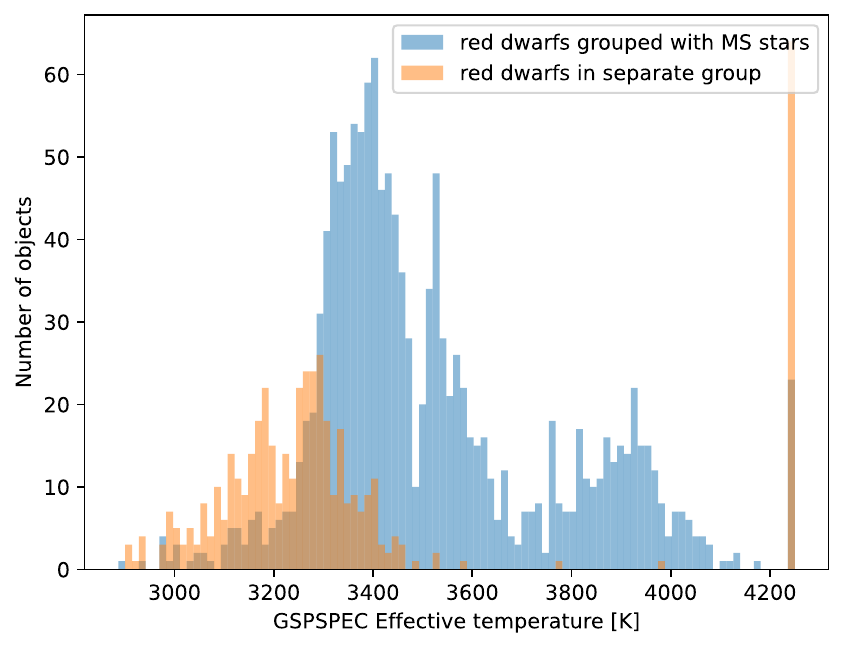}
    \caption{Distribution of best-fit slope and GSPSPEC $T_{\rm eff}$ for the Red dwarf groups. A division is visible between flat-continuum and sloped spectra. (a) Red dwarfs – distributions of fit slopes. (b) Red dwarfs – distributions of GSPSPEC $T_{\rm eff}$. The isolated bin near $4250$~K in the GSPSPEC panel is not interpreted as a distinct physical temperature population; this value matches the fixed $T_{\rm eff}=4250\pm500$~K correction documented for problematic K/M-type giant GSPSPEC parametrisations \citep[Section 8.6]{Recio_Blanco_2023}.}
    \label{fig:red_dwarf_histograms}
\end{figure}

Figure \ref{fig:red_dwarf_hr_umap} shows the median RVS spectrum for each group. In the first group, we observe a relatively flat continuum, while the second group has a noticeable slope. This is in conflict with our expectation that the RVS spectra were continuum-normalized. Moreover, there is a difference in the GSPSPEC-derived effective temperature ($T_{eff}$) values \citep{Recio_Blanco_2023} between these two groups. The first group has a median $T_{eff}$ of 3,447K, while the second group is cooler with a median $T_{eff}$ of 3,264K (Figure \ref{fig:red_dwarf_histograms}, right).

The distinction between the two groups becomes apparent when we examine their slope distributions. By fitting linear models to the spectra using the L1 norm as the optimization criterion, we measure the slope of all these spectra. The histograms of slope values (Figure \ref{fig:red_dwarf_histograms}, left) show distinct non-overlapping distributions. Clearly the RF picks up on a true separation in the data. Further investigation reveals that indeed the RVS spectra are continuum-normalized by the Gaia pipeline, but only if their matched spectral template's effective temperature is greater than $3500$~K \citep{gaia_docs}. This is likely the root of this artificial discontinuity in the similarity of the spectra for these stars (Figure \ref{fig:red_dwarf_histograms}).

\subsubsection{Red supergiants}
\begin{figure*}
  \centering
    \includegraphics[width=\textwidth]{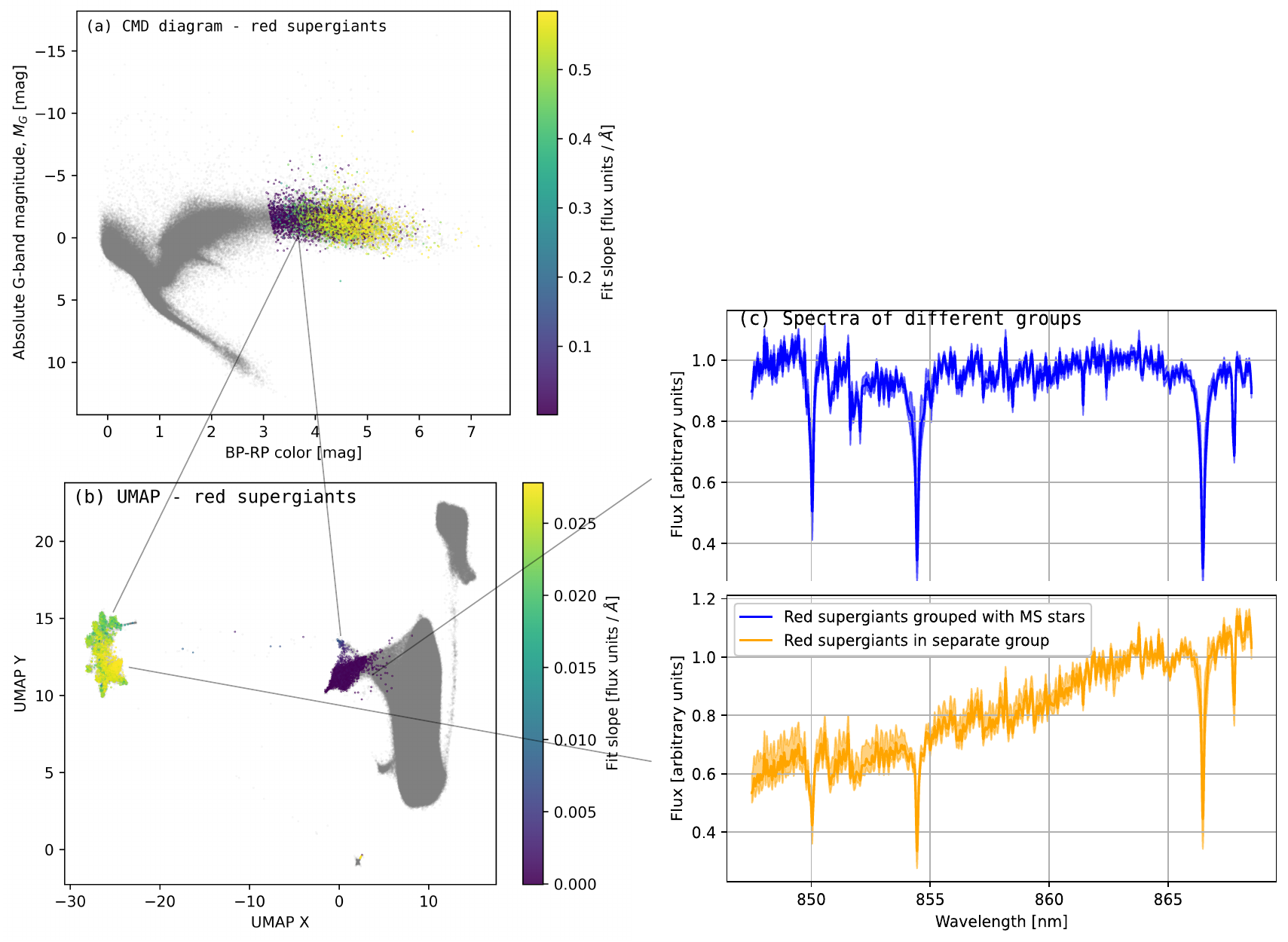}
    \caption{CMD-selected luminous cool-star region highlighted in CMD and UMAP together with the median spectra of its two UMAP groups. The selection uses absolute G-band magnitude ($M_G<5$) and color ($BP-RP>3$). We retain the informal ``Red supergiants'' label in the text as shorthand only; this photometric cut is not expected to isolate a pure physical RSG sample.}
    \label{fig:red_supergiant_hr_umap}
\end{figure*}

The Red supergiants, referring here only loosely to stars in the observed CMD region with a G-band absolute magnitude smaller than 5 and $BP-RP$ color greater than 3, seem to also separate into two groups, as can be seen in Figure \ref{fig:red_supergiant_hr_umap}. We stress that this label is only a shorthand for a luminous cool-star locus in the CMD, not a claim that all selected objects are bona fide red supergiants. A cut this broad can also include other cool luminous stars, and robust RSG identification generally requires stronger luminosity/temperature constraints together with extinction treatment \citep{Healy2024RSG}. About half of these 20,000 stars are situated near the main sequence stars, while the other half is in a separate group. This is a much more balanced division than what we have seen for the Hot stars and the Red dwarfs. 
Because this selection is made in the observed CMD (without dereddening), interstellar extinction can shift stars in both color and magnitude and may sharpen or blur apparent CMD discontinuities. We therefore interpret this cut as a practical selection in observed photometric space, not as an extinction-corrected physical boundary.

Figure \ref{fig:red_supergiant_hr_umap}c shows the median RVS spectra for each group. One group exhibits a relatively flat continuum, while the separate group displays a noticeable slope, again in conflict with our expectation that the RVS spectra are continuum-normalized.

\begin{figure}
    \centering 
        \includegraphics[width=0.48\linewidth]{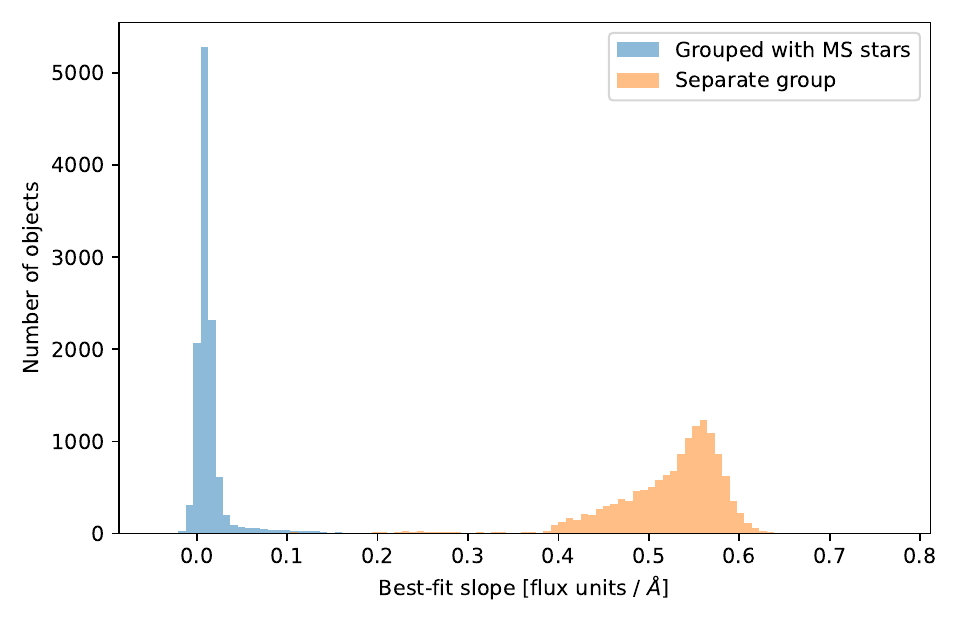}  
    \hfill
        \includegraphics[width=0.48\linewidth]{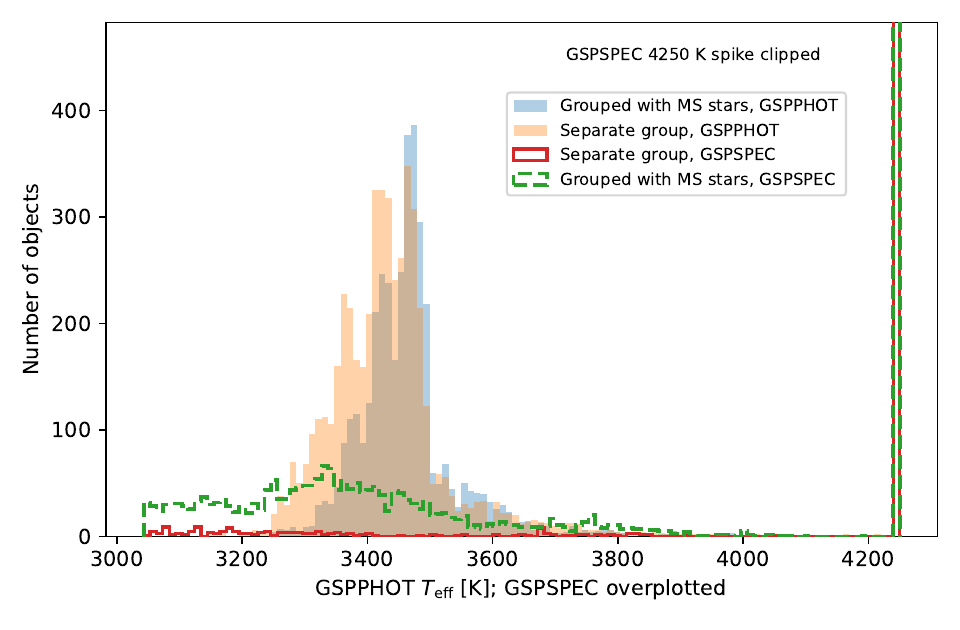}
    \caption{Distribution of best-fit slope (left) and GSPPHOT $T_{\rm eff}$ (right, filled histograms) for the two groups within this CMD-selected luminous cool-star region. GSPSPEC $T_{\rm eff}$ values are overplotted as step histograms in the right panel. A separation between flat-continuum and sloped spectra is visible, while neither temperature product shows a comparably clean division. Its origin is therefore likely a processing artifact tied to continuum shape.}
    \label{fig:red_supergiant_slope_teff_histograms}
\end{figure}

The effective temperature ($T_{eff}$) values obtained from the GSPSPEC pipeline \citep{Recio_Blanco_2023} do not provide conclusive information for these stars; many objects either lack a solution or cluster near $4250$~K. This value matches the fixed $T_{\rm eff}=4250\pm500$~K correction documented for problematic K/M-type giant GSPSPEC parametrisations \citep[Section 8.6]{Recio_Blanco_2023}, so we do not interpret the pile-up as evidence for a meaningful recovered stellar temperature. Figure \ref{fig:red_supergiant_slope_teff_histograms} shows photometric GSPPHOT $T_{\rm eff}$ values as filled histograms and GSPSPEC $T_{\rm eff}$ values as step histograms. The GSPSPEC overlay does not show a clean physical temperature boundary for the two groups, while the GSPPHOT distributions are broad and strongly overlapping. Thus the displayed evidence for the split is instead the striking bimodal distribution of continuum slope values, which closely matches the machine-learned groups. This is consistent with the same continuum-normalization issue discussed above for cooler templates: the documented $3500$~K threshold applies to the RVS template used by the Gaia processing, whereas GSPPHOT and GSPSPEC $T_{\rm eff}$ are later catalogue products and need not reproduce that threshold as a sharp boundary in this figure. We therefore interpret the split as a likely processing artifact tied to continuum shape, not as a separate physical temperature sequence.
This interpretation is also consistent with extinction primarily affecting the CMD placement, while the RVS-window similarity metric is expected to be less sensitive to reddening itself, except indirectly through lower SNR and possible diffuse interstellar band signatures.

\Needspace{12\baselineskip}
\subsection{Weirdness score and outlier detection}
\Needspace{10\baselineskip}
\subsubsection{Weirdness score spike}\label{sec:weirdness_spike}
\begin{figure}
    \centering
         \includegraphics[width=0.48\linewidth]{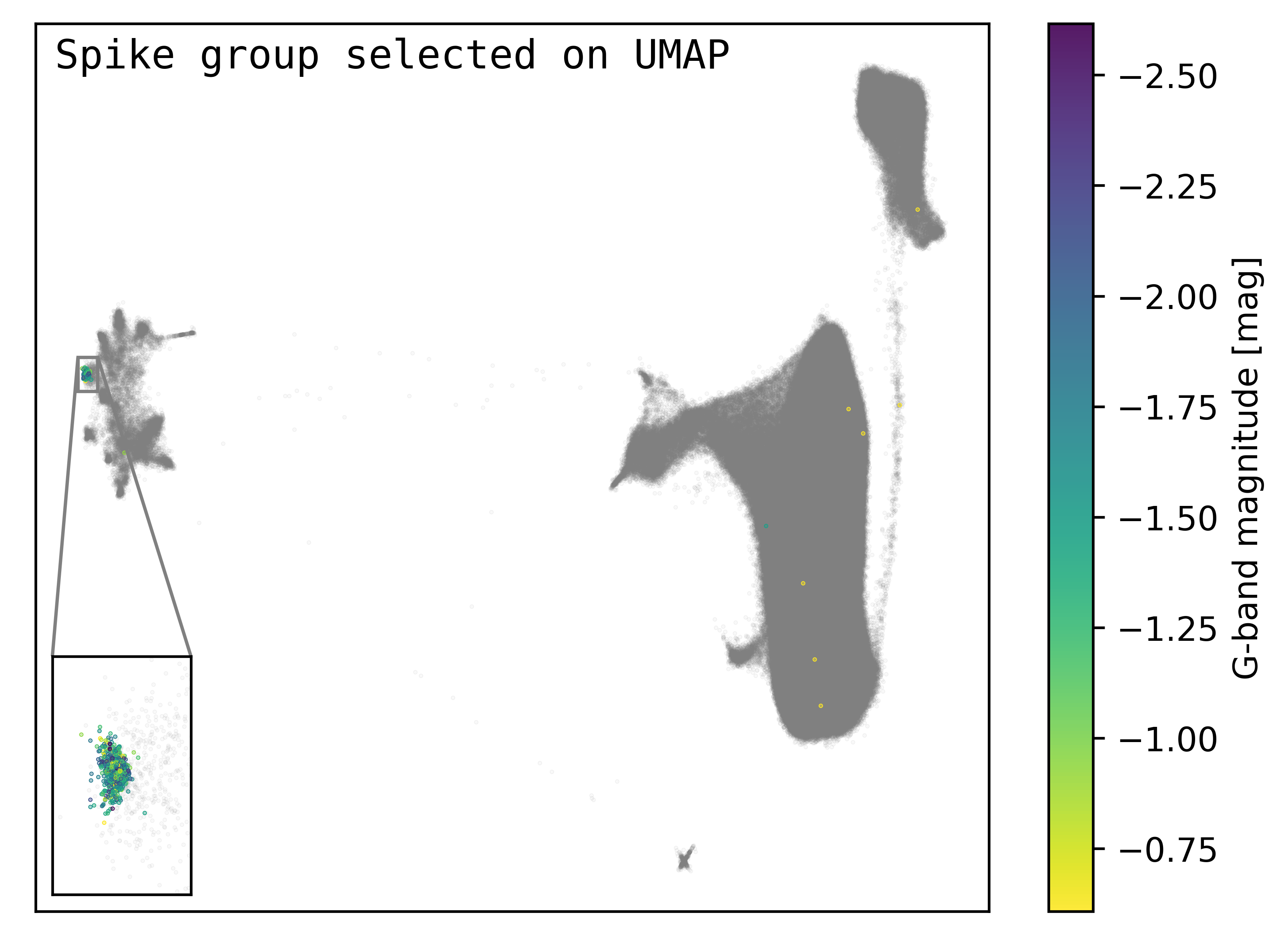}
    \hfill
        \includegraphics[width=0.48\linewidth]{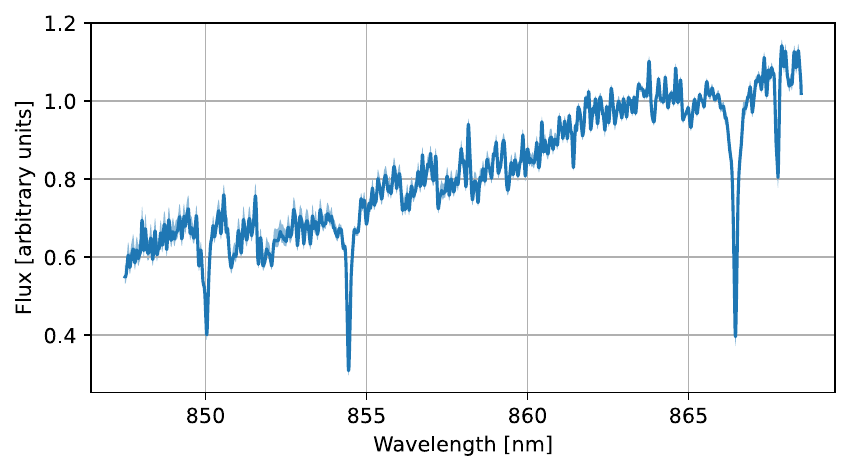}
       \caption[Objects found within the weirdness score spike bin]{Objects found within the weirdness score spike bin. On the left we show their location in the UMAP representation, where most of these stars from the CMD-selected luminous cool-star region appear clumped together. On the right we show their median RVS spectrum and the very small deviation around it.}
    \label{fig:weirdness_spike}
\end{figure}

In Figure \ref{fig:weirdness_hist}, we observe a conspicuous spike in the histogram of weirdness scores, consisting of approximately 500 additional objects clustering around a value of 0.76. We find that this is due to 474 stars from this same CMD-selected luminous cool-star region that have the exact same weirdness score. They cluster tightly in UMAP (see Figure \ref{fig:weirdness_spike}), and have, as one would expect, nearly identical spectra, also shown. Because of that they consistently map to the same terminal leaves in the RF classifier, and are indistinguishable in our space.

\Needspace{10\baselineskip}
\subsubsection{The weirdest stars in Gaia RVS}\label{sec:weirdest_objects}

As another demonstration for the potential uses of our tools, we briefly analyze here the weirdest stars in Gaia RVS, according to our metric. We expect the weirdness scores to represent an object's level of oddity (hence the name). Our aim here is not to perform an exhaustive search for outliers, but rather to validate the usefulness of this representation. 

In table \ref{tab:weirdest_table} we present the Top-20 weirdness sample together with each object's designation (if we find any) and our impressions on their spectral peculiarities. We formed our impressions by visually inspecting each spectrum together with the median spectra of its 100 nearest neighbors in the high-dimensional embedding. This allows us to hypothesize on what made that specific object stand out.
For this small illustrative sample, the SIMBAD cross-reference was verified manually on an object-by-object basis rather than through a blind bulk match with a fixed acceptance radius. We used the Gaia DR3 source positions/identifiers to inspect the nearest SIMBAD counterpart in real time. We also inspected the sky-image context for each object and checked the local search vicinity for competing candidate stars; no such competing counterpart was found. In all 20 cases the association was unambiguous and showed no nearby-source confusion, so the corresponding SIMBAD designation was used as contextual information to check whether the object had a known classification or relevant literature.

\begin{table}
    \centering
    \renewcommand{\arraystretch}{1} 
    \begin{tabular}{c|l|l|l|l} 
        \hline
        \textbf{Num} & \textbf{Gaia DR3 ID} & \textbf{Known SIMBAD Type} & \textbf{Assigned group} & 
        \textbf{Comments and comparison} \\[-0.5ex]
        & & & & \textbf{to most similar spectra} \\
        \hline
        1 & 4296238438875177600 & Mira & Mira & Ca II in emission \\
        2 & 3453437075044849408 & S star & S star & step function in spectrum \\
        3 & 2808855804962475776 & Mira & Mira & missing Ca II lines \\
        4 & 160026083402696448 & Mira & Mira & weak Ca II lines \\
        5 & 4657630731032546176 & Yellow supergiant & Yellow supergiant & deep and broad Ca II lines \\
        6 & 6654541175020312704 & -- & Unclassified & No clear distinguishing feature in visual comparison \\
        7 & 4662293897281149824 & Blue supergiant in LMC & Blue supergiant & broad second Ca II line \\
        8 & 6746809304604808320 & -- & Unclassified & No clear distinguishing feature in visual comparison \\
        9 & 4066740537925096064 & Yellow-white supergiant & Yellow supergiant & deep and broad Ca II lines \\
        10 & 5305186798527480064 & -- & Unclassified & No clear distinguishing feature in visual comparison \\
        11 & 4049882207145302912 & -- & Unclassified & Broader Ca II lines \\
        12 & 4659451968980169984 & Yellow supergiant & Yellow supergiant & Continuum differences \\
        13 & 5264710854914674944 & -- & Unclassified & No clear distinguishing feature in visual comparison \\
        14 & 4318129195634230912 & S star & S star & step function in spectrum \\
        15 & 5342074481609304832 & -- & Unclassified & No clear distinguishing feature in visual comparison \\
        16 & 5313996871648315904 & S star & S star & Step function \\
        17 & 6363620500071779584 & Cool carbon star & Cool carbon star & Line-rich spectrum \\
        18 & 6162222229360137344 & Cool carbon star & Cool carbon star & Line-rich spectrum \\
        19 & 5235055239624830336 & S star & S star & Step function, weak Ca II lines \\
        20 & 4660251348288638848 & Classical Cepheid Variable & Cepheid & Lacking some lines \\
        \hline
    \end{tabular}
    \caption{Top-20 weirdness sample (highest weirdness-score stars), with SIMBAD types assigned from manually verified counterparts, an explicit per-object assigned-group label, and brief notes relative to the 100 nearest neighbors in the RF embedding. All 20 SIMBAD associations were unambiguous.}
    \label{tab:weirdest_table}
\end{table}

From table \ref{tab:weirdest_table}, it is clear that a significant portion of the objects in the Top-20 weirdness sample are known to be rather unusual stars. These include Mira variables, S stars, stars in the LMC, cool carbon stars, and Cepheids. This is further evidence that our distance matrix captures true properties of the sample, and that the weirdness score traces true rarity.

\begin{figure}[!htbp]
    \centering
    \includegraphics[width=0.48\textwidth]{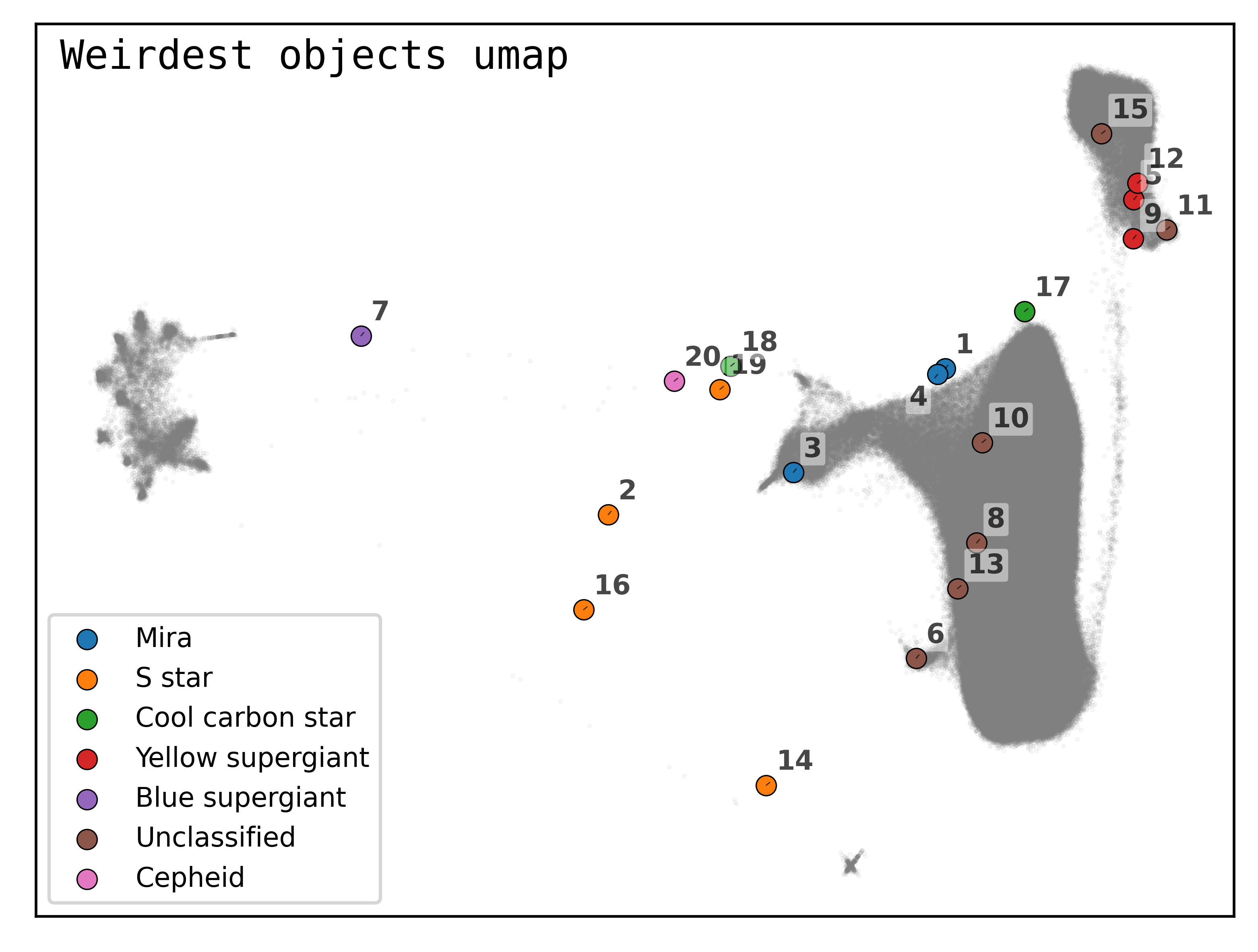}
    \hfill
    \includegraphics[width=0.48\textwidth]{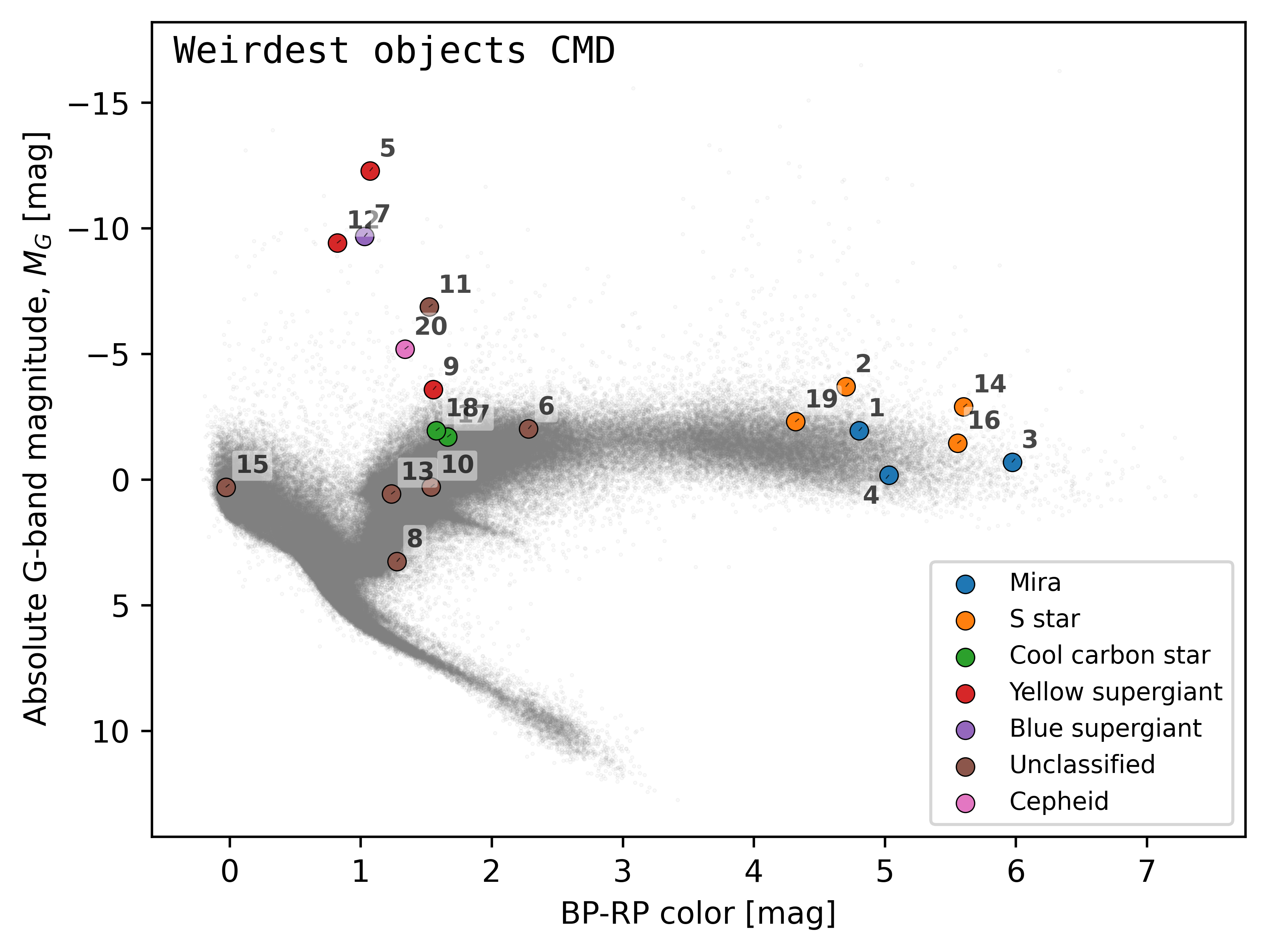}
    \caption{Top-20 weirdness sample shown in the UMAP representation (left) and the CMD (right). The object numbering follows Table \ref{tab:weirdest_table}.}
    \label{fig:weirdest_umap}
\end{figure}

Figure \ref{fig:weirdest_umap} shows the locations of these objects within the UMAP representation. While a large percentage of them appear in low density areas as would be expected, some of them reside within dense groups in the UMAP representation, pointing to the fact that some of the aspects of the high-dimensional structure cannot entirely be preserved in the dimensionality reduction process, as can be expected. This is also expected from the definitions: weirdness is a global score (the mean distance from one object to all others), while UMAP is optimized to preserve local neighborhoods (set here by $num\_neighbors=25$). Therefore a star can be globally unusual in the 500D RF embedding yet still lie inside a locally dense UMAP region. Consequently, the analysis in this subsection should be interpreted as a study of globally unusual objects under one specific outlier definition, not as a complete census of local anomalies. A local method such as LOF (with tunable neighborhood size $k$) would emphasize a different notion of rarity and may reprioritize objects in dense regions. A more complete search for outliers, even just with this single distance metric, should be executed systematically not spectrum by spectrum, but using the UMAP representation to analyze them in groups, as done, for example, by \citet{10.1093/mnras/sty348}. We now examine in more detail the types of weird stars that we find.

\begin{figure}[!htbp]
    \centering
    \begin{minipage}[t]{0.31\textwidth}
        \centering
        \includegraphics[width=\textwidth]{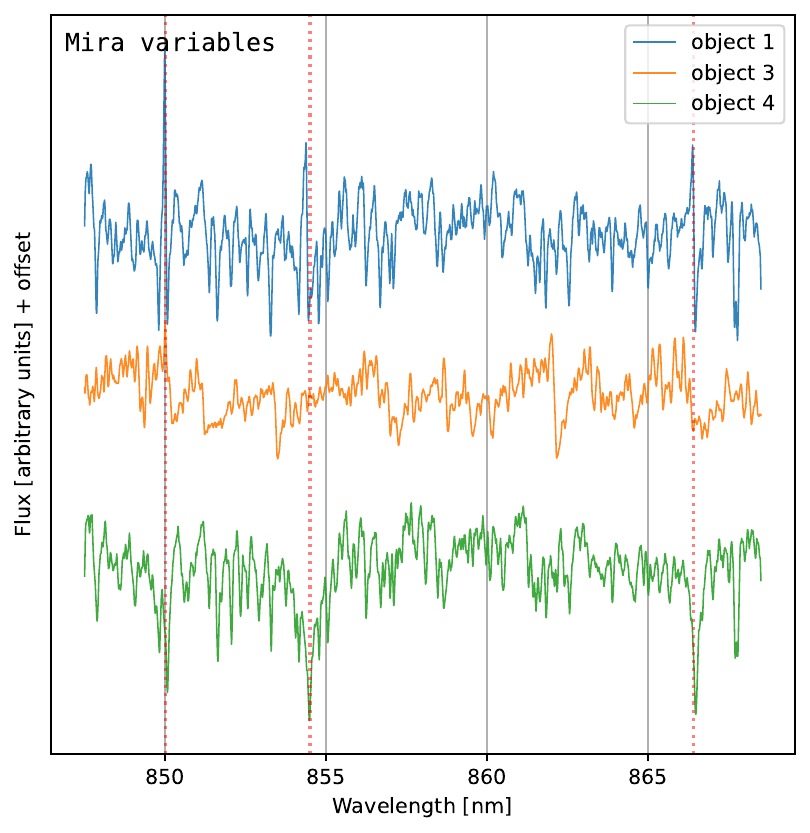}
        \caption{Top-20 weirdness sample: Mira variables. Red dotted lines indicate the location of the Ca II lines, which show weakened absorption, or even emission.}
        \label{fig:weird_mira}
    \end{minipage}\hfill
    \begin{minipage}[t]{0.31\textwidth}
        \centering
        \includegraphics[width=\textwidth]{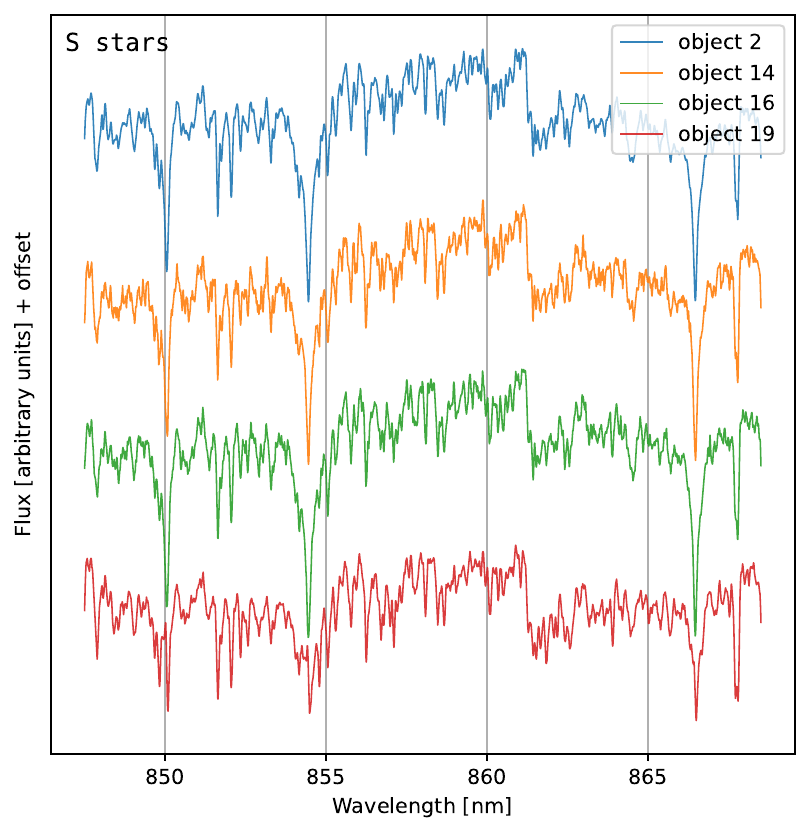}
        \caption{Top-20 weirdness sample: S stars with a broad absorption trough near $862\,\text{nm}$. This is probably due to VO absorption.}
        \label{fig:weird_s}
    \end{minipage}\hfill
    \begin{minipage}[t]{0.31\textwidth}
        \centering
        \includegraphics[width=\textwidth]{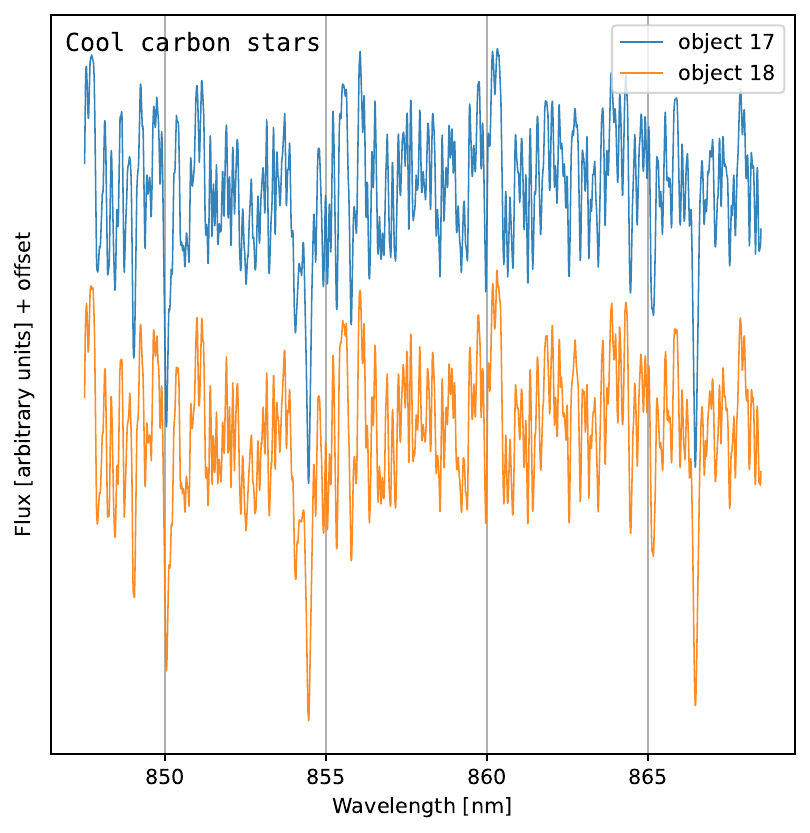}
        \caption{Top-20 weirdness sample: Cool carbon stars and their busy spectra.}
        \label{fig:weird_cool_carbon}
    \end{minipage}

    \vspace{0.5em}

    \begin{minipage}[t]{0.31\textwidth}
    \centering
        \includegraphics[width=\textwidth]{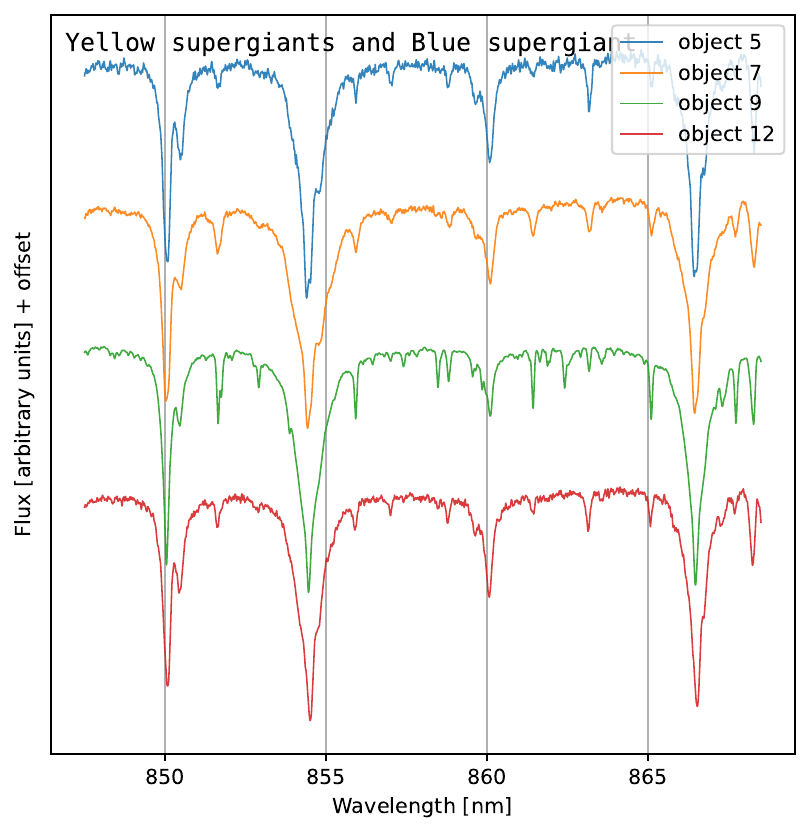}
        \caption{Top-20 weirdness sample: Yellow supergiants (objects 5, 9 and 12) and the Blue supergiant (object 7).}
        \label{fig:weird_yellow}
    \end{minipage}\hfill
    \begin{minipage}[t]{0.31\textwidth}
        \centering
        \includegraphics[width=\textwidth]{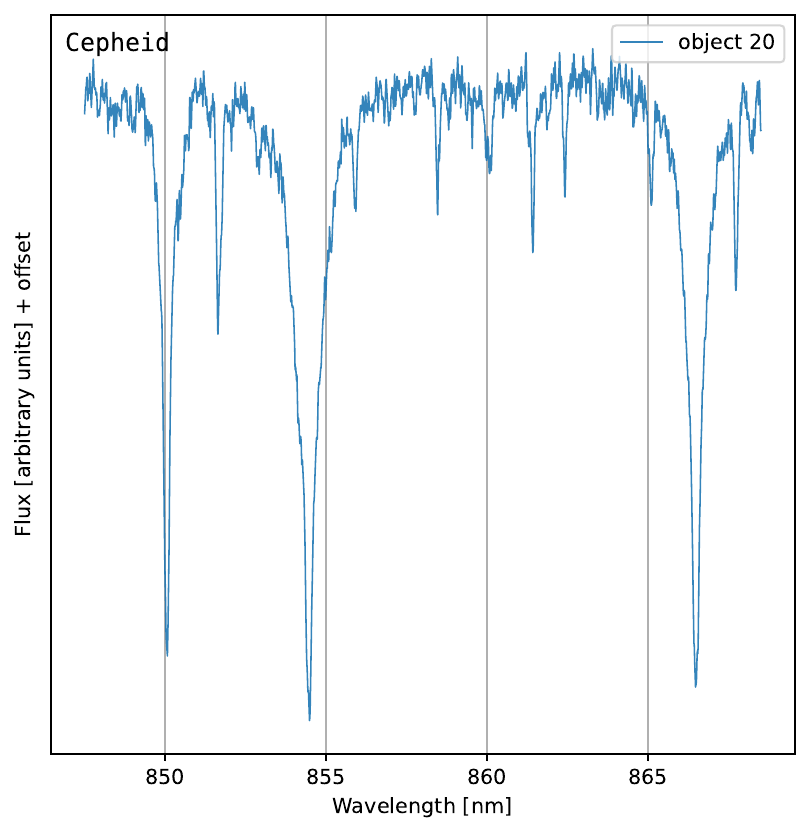}
        \caption{Top-20 weirdness sample: Cepheid variable star, lacking some lines.}
        \label{figure:cepheid_variable}
    \end{minipage}\hfill
    \begin{minipage}[t]{0.31\textwidth}
        \centering
        \includegraphics[width=\textwidth]{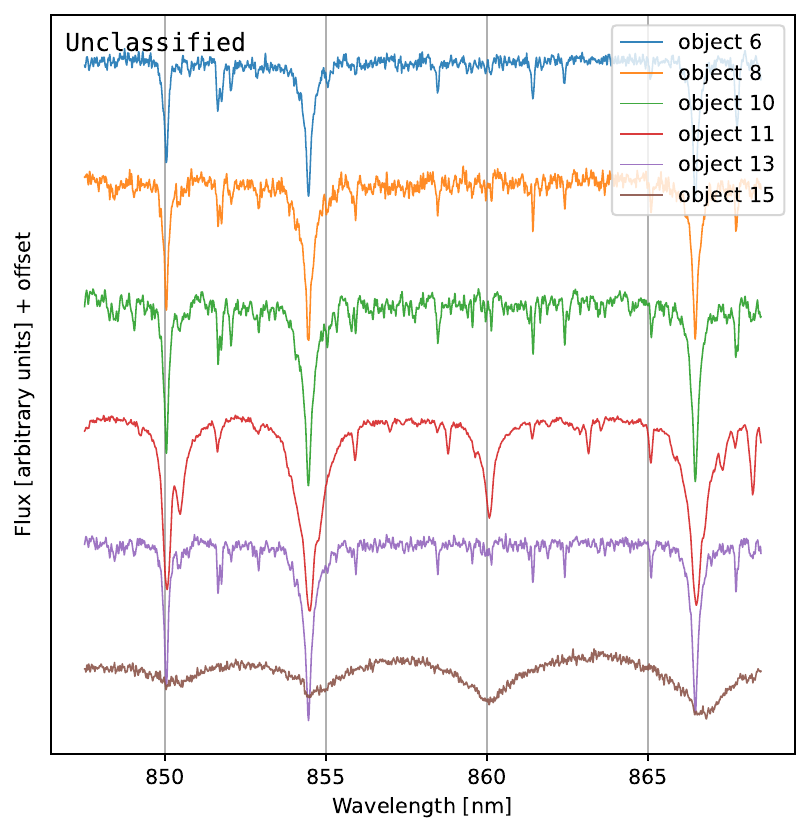}
        \caption{Top-20 weirdness sample: Unclassified objects. For all but object 11 the nearest-neighbor median is virtually identical to the outlier; object 11 clearly differs from its neighbors.}
        \label{fig:weird_unknown}
    \end{minipage}
\end{figure}

Mira variables are a type of pulsating variable star characterized by their large amplitudes over relatively long periods, of over a year. These stars are typically Red giants or supergiants in later stages of stellar evolution, with extended atmospheres and sometimes circumstellar dust clouds \citep{Hofner2018-ay}. The Mira variables in our Top-20 weirdness sample have conspicuously unusual Ca II lines compared to their 100 nearest neighbors. Figure \ref{fig:weird_mira} shows the spectra of the three identified Mira stars, together with the median spectrum and variance of their 100 nearest neighbors. These variations in the Ca II lines are a known behavior in Mira variable stars and can be attributed to fluorescence from $H_\epsilon$ transitions \citep{Castelaz_2000}, VO/TiO blanketing, or chromospheric emission \citep{0710.1491}.
In the UMAP representation (Figure \ref{fig:weirdest_umap}), these stars fall at the edges of the giant and supergiant groupings. Star 1 and 4 lie close together, while 3 is apart. Spectrally, 1 and 4 differ in their Ca II lines, but indeed, the rest of their spectral features appear similar. Star 3 differs from them at every wavelength. Analysis of the embedding confirms that stars 1 and 4 are nearest neighbors, but 3 is not their neighbor. This implies that even within the restricted wavelength range of the RVS spectra, a rare class of stars like Mira variables can have enough variance in its spectral features, and enough stars in the sample, in order to find these variations directly from the data. While these objects are all Miras according to SIMBAD, we identify at least two distinct spectral behaviors.

S stars are characterized by having an excess of s-process elements in their spectra, particularly elements like strontium, yttrium, zirconium, and barium. These stars are believed to be transition objects between M-type giants and carbon stars on the asymptotic giant branch \citep{stype}. Many S stars are also classified as Mira variables, with the S-star classification originally being a subclass of Mira variable stars \citep{1922ApJ....56..457M}. Figure \ref{fig:weird_s} shows the spectra of the four identified S stars together with the median spectrum and variance of their 100 nearest neighbors. All seem to have a broad absorption trough near $862 nm$, compared to a much weaker feature in their nearest neighbors. This is probably due to VO absorption \citep{1997AJ....114.1584C} as can be seen at other wavelengths with similar molecules such as ZrO, LaO, and TiO \citep{1954ApJ...120..484K}. 
In the UMAP representation (Figure \ref{fig:weirdest_umap}), these objects fall completely outside any grouping, in the space between the two supergiant groupings. These objects appear spread apart even though their spectra seem visually similar. Analysis of the embedding confirms that objects 2, 14, and 16 are in-fact each-others' nearest neighbors. Object 19 is farther away in the embedding, possibly owing to its weaker Ca II absorption.

Carbon stars have prominent carbon features, due to carbon overabundance. These are typically evolved Red giants with circumstellar shells, soot, disks, or clouds \citep{carbon}. The cool carbon stars we found emit a line-rich spectrum compared to the rest of the dataset and their neighbors, some lines rivaling the strength of the Ca II triplet. This is a known behavior of carbon stars, and the lines are attributed to CN absorption \citep{2003ASPC..298..451P}. Figure \ref{fig:weird_cool_carbon} shows the spectra of the two identified cool carbon stars. In the UMAP representation (Figure \ref{fig:weirdest_umap}), these stars fall at the edges of the giant and supergiant groupings. Analysis of the embedding confirms that they are in-fact each-others' nearest neighbor despite the distance between them in the UMAP representation.

Our Top-20 weirdness sample includes 3 objects assigned to the Yellow supergiant group (objects 5, 9 and 12 in Table \ref{tab:weirdest_table}) and one object assigned to the Blue supergiant group (object 7, located in the LMC). Figure \ref{fig:weird_yellow} shows the strong and broad Ca lines that one expects. In the UMAP representation (Figure \ref{fig:weirdest_umap}), the yellow supergiants fall within the Blue supergiant grouping, while the Blue supergiant in the LMC falls outside any grouping. The latter fact is surprising given that visually object 7 appears very similar to other Blue supergiants in the grouping, and specifically to objects 5, 9 and 12 as well. Analysis of the embedding confirms that these objects are each other's nearest neighbors even though object 7 is located farther away in the UMAP representation. Comparison to spectra of 100 nearest neighbors shows that these objects have even broader lines than typical. We measured the equivalent widths of the three lines and found an average for supergiants of $0.37nm$, $0.78nm$ and $0.53nm$ for the $850nm$, $854.5nm$ and $866.4nm$ lines respectively, while for the 100 nearest neighbors those average equivalent widths are $0.23nm$, $0.46nm$, and $0.40nm$.

The lone classical Cepheid in our Top-20 weirdness sample (Table \ref{tab:weirdest_table}, object 20) will now be discussed. Cepheids are luminous, pulsating F–G (sometimes K) supergiants whose effective temperature, surface gravity and line profiles change substantially over a single pulsation cycle; consequently their spectra can appear depleted or enhanced in specific features depending on phase \citep{pietrukowicz_soszyński_udalski_2021, Anderson_2016}. The Cepheid flagged by our metric shows a noticeable absence (or weakening) of some lines within the RVS window compared to its 100 nearest neighbors, e.g at $852nm$ and $855nm$ — a behaviour that can naturally arise from phase-dependent temperature changes or shock-induced line-profile variability.

Figure \ref{fig:weird_unknown} shows the spectra of the six objects that remain unclassified in SIMBAD and are therefore assigned the Unclassified group label in Table \ref{tab:weirdest_table}. In the UMAP representation (Figure \ref{fig:weirdest_umap}) these objects fall within known broader groupings -- object 6 in the Red dwarf group, objects 8, 10, 13 in the main sequence and giant group, and objects 11, 15 in the Blue supergiant group. Their spectra are visually similar to that of their neighbors, which does not contradict their high weirdness scores because weirdness reflects global, not local, dissimilarity. Object 11 is the only case where the oddity is also easily apparent in local spectral comparison -- its CaII lines are noticeably broader compared to its nearest neighbors.

\clearpage

%% file: 07_discussion.tex
\section{Summary \& Conclusion}
\label{s:discussion}
S-Disco, the Gaia stellar spectra and characteristics exploration tool which is presented in this paper allows one 
to explore Gaia RVS spectra using embeddings derived from self-supervised metric learning. We combine RF classification for metric learning and UMAP for dimensionality reduction, allowing the identification of groups and anomalies within the dataset. 
Our analysis processes 312,295 spectra out of the 997,162 Gaia DR3 RVS spectra by selecting the SNR $>50$ subset, specifically to limit noise-driven false anomalies and prioritize spectra whose outlier status is physically interpretable.

UMAP projections, while useful for visualizing the data structure, inherently involve some information loss. As expected, this information loss is greatest in low-density areas of the embedding, which are correspondingly the areas with high-weirdness objects. This is evident from the occasional discrepancy between the UMAP representation and the high-dimensional embedding, as seen with the placement of certain high-weirdness score objects. 
The interpretation of outliers should therefore distinguish between global and local behavior: our weirdness score is a global statistic (average distance to the full sample), whereas UMAP emphasizes local neighborhood structure. As a result, some globally weird objects can still appear inside dense local regions in the 2D projection.
More generally, outlier rankings depend on the scoring definition. In particular, local-density methods such as LOF provide a complementary, scale-controlled view through the choice of neighborhood size $k$ (small $k$ for local anomalies, larger $k$ for broader-context anomalies). We do not include an alternative score here, but this is a natural extension for future versions of the analysis and platform.

Our work demonstrates that with such methods, one can learn a meaningful similarity metric for the RVS spectra, despite their narrow wavelength range. We identify various underlying patterns and structures within the data without supervised training. This includes a small transitory group between the main sequence and Hot stars, groups of nearly identical stars, and subtle systematics that are due to the way continuum subtraction was applied by Gaia. This shows the potential of metric learning for the important, yet mundane tasks of data cleaning and debugging. Perhaps more interestingly, most of the top 20 weirdest objects we study are previously known unusual stars, such as Mira variables, S stars, and cool carbon stars. In the context of recent contrastive and foundation-model work on stellar and other astronomical spectra, our results support the same general picture that representation learning can recover physically meaningful organization from largely unlabeled survey data, while also illustrating that a comparatively lightweight RF-based metric remains scientifically useful and operationally simple for survey-scale exploration.

This tool enables the community to study this sample in a novel way, and uncover insights hidden in this large and unique dataset. While we show the substantial capabilities that our method provides, we made specific hyperparameter choices, in both RF and UMAP, that impact the results. Although our selection is guided by empirical testing and qualitative expectations, it remains inherently subjective. For UMAP in particular, one parameter setting cannot represent all scales equally, and alternative settings may expose additional physically meaningful substructure that is compressed in our default map. We plan to enrich the platform in the future, to include other possibly useful mappings (perhaps following Gaia DR4). 
Likewise, CMD-defined selections in this work are based on observed (not dereddened) photometry, so extinction can shift objects across heuristic CMD boundaries; this should be kept in mind when interpreting discontinuities seen in CMD space.